\documentclass[
reprint,
superscriptaddress,
 amsmath,amssymb,
 prl,
]{revtex4-2}

\usepackage{graphicx}
\usepackage{dcolumn}
\usepackage{bm}
\usepackage{hyperref}
\hypersetup{
    colorlinks=true,
    allcolors=blue
}

\usepackage[final]{changes} 

\definechangesauthor[name={Maximilian Zanner}, color=red]{max}
\definechangesauthor[name={Romain Albert}, color=red]{rom}
\definechangesauthor[name={Gerhard Kirchmair}, color=red]{ger}

\usepackage[mathlines]{lineno}
\usepackage{siunitx}
\DeclareSIUnit\bar{bar}
\usepackage{csquotes}
\usepackage{braket}
\usepackage{wasysym} 
\usepackage{todonotes}

 \usepackage{color}

\usepackage[protrusion=true, expansion=true]{microtype}

\newcommand{\beginsupplement}{%
        \setcounter{table}{0}
        \renewcommand{\thetable}{S\arabic{table}}%
        \setcounter{figure}{0}
        \renewcommand{\thefigure}{S\arabic{figure}}%
        \setcounter{equation}{0}
     }
     
\newcommand{\fakesection}[1]{%
  \par\refstepcounter{section}
  \sectionmark{#1}
  \addcontentsline{toc}{section}{\protect\numberline{\thesection}#1}
}

\begin{document}

\title{Spatial Addressing of Qubits in a Dispersive Waveguide}


\author{Maximilian Zanner}
\email{These two authors contributed equally\\zanner.maximilian@gmail.com}
\affiliation{Institute for Quantum Optics and Quantum Information, Austrian Academy of Sciences, 6020 Innsbruck, Austria}
\affiliation{Institute for Experimental Physics, University of Innsbruck, 6020 Innsbruck, Austria}

\author{Romain Albert}
\email{These two authors contributed equally\\romain@silent-waves.com} 
\affiliation{Institute for Quantum Optics and Quantum Information, Austrian Academy of Sciences, 6020 Innsbruck, Austria}
\affiliation{Institute for Experimental Physics, University of Innsbruck, 6020 Innsbruck, Austria}
\affiliation{Silent Waves, 25 avenue des Martyrs, 38042 Grenoble cedex 09, France}

\author{Eric I. Rosenthal}
\affiliation{E. L. Ginzton Laboratory, Stanford University, Stanford, California 94305, USA}

\author{Silvia Casulleras}
\affiliation{Institute for Quantum Optics and Quantum Information, Austrian Academy of Sciences, 6020 Innsbruck, Austria}
\affiliation{Institute for Theoretical Physics, University of Innsbruck, 6020 Innsbruck, Austria}

\author{Ian Yang}
\affiliation{Institute for Quantum Optics and Quantum Information, Austrian Academy of Sciences, 6020 Innsbruck, Austria}
\affiliation{Institute for Experimental Physics, University of Innsbruck, 6020 Innsbruck, Austria}

\author{Christian M. F. Schneider}
\affiliation{Institute for Quantum Optics and Quantum Information, Austrian Academy of Sciences, 6020 Innsbruck, Austria}
\affiliation{Institute for Experimental Physics, University of Innsbruck, 6020 Innsbruck, Austria}

\author{Oriol Romero-Isart}
\affiliation{Institute for Quantum Optics and Quantum Information, Austrian Academy of Sciences, 6020 Innsbruck, Austria}
\affiliation{Institute for Theoretical Physics, University of Innsbruck, 6020 Innsbruck, Austria}
\affiliation{ICFO - Institut de Ciencies Fotoniques, The Barcelona Institute of Science and Technology, 08860 Castelldefels (Barcelona), Spain}
\affiliation{ICREA, Passeig Lluis Companys 23, 08010, Barcelona, Spain}

\author{Gerhard Kirchmair}
\affiliation{Institute for Quantum Optics and Quantum Information, Austrian Academy of Sciences, 6020 Innsbruck, Austria}
\affiliation{Institute for Experimental Physics, University of Innsbruck, 6020 Innsbruck, Austria}

\date{\today}


\begin{abstract}
Waveguide quantum electrodynamics -- the study of atomic systems interacting with propagating electromagnetic fields -- is a powerful platform for understanding the complex interplay between light and matter. Qubit control is an indispensable tool in this field, and most experiments have so far focused on narrowband electromagnetic waves that interact with qubits at specific frequencies. This interaction, however, changes significantly with fast, broadband pulses, as waveguide properties like dispersion affect the pulse evolution and its impact on the qubit. Here, we use dispersion to achieve spatial addressing of superconducting qubits separated by a sub-wavelength distance within a microwave waveguide. This novel approach relies on a self-focusing effect to create a position-dependent interaction between the pulse and the qubits. This experiment emphasizes the importance of dispersion in the design and analysis of quantum experiments, and offers new avenues for the rapid control of quantum states.

\end{abstract}

\maketitle



A propagating wave is said to disperse if its shape changes over time. Dispersion arises either due to the material properties of the medium~\cite{brabecIntenseFewcycleLaser2000a} or from the geometric confinement of the propagating field~\cite{pozarMicrowaveEngineering2012, joannopoulosPhotonicCrystalsPutting1997c}. Understanding the propagation of electromagnetic waves in different media, including the effects of dispersion, plays a central role in modern information technology. This understanding is essential both in the classical regime, such as in telecommunications (e.g. fiber optic cables), and in the quantum regime, where single photons must be routed and controlled with high fidelity. Consequently, dispersion is highly relevant to the field of waveguide quantum electrodynamics (QED), which investigates fundamental light-matter interactions and quantum protocols involving propagating waves~\cite{sheremetWaveguideQuantumElectrodynamics2023, massonAtomicwaveguideQuantumElectrodynamics2020}. 

Waveguide QED using superconducting circuits has been successful in investigating quantum phenomena where quantum emitters interact with a propagating microwave field. Landmark experiments include the first realization of strong coupling between a superconducting qubit and a one-dimensional transmission line~\cite{astafievResonanceFluorescenceSingle2010b}, the creation of super- and subradiant states formed by multiple emitters~\cite{looPhotonMediatedInteractionsDistant2013a,zannerCoherentControlMultiqubit2022a}, the demonstration of single photon routers~\cite{hoiDemonstrationSinglePhotonRouter2011}, and the achievement of multi-photon entangled states~\cite{kannanGeneratingSpatiallyEntangled2020a} as well as interacting bound states~\cite{sundaresanInteractingQubitPhotonBound2019}. Other significant achievements include entangling distant quantum systems to perform Bell tests~\cite{storzLoopholefreeBellInequality2023}, creating giant atoms using self-interference~\cite{kannanWaveguideQuantumElectrodynamics2020} and realizing chiral interactions and uni-directional photon emitters~\cite{rosariohamannNonreciprocityRealizedQuantum2018, kannanOndemandDirectionalMicrowave2023, joshiResonanceFluorescenceChiral2023}. In all these experiments, quantum emitters interact with waveguide photons within a narrow bandwidth, allowing the impact of dispersion to be neglected.
 
However, dispersion significantly affects the temporal shape of short, wideband wavepackets in quantum communication protocols, similar to its impact in classical communication technology. Consequently, understanding wavepacket evolution due to dispersion and appropriately tailoring control pulses will be crucial for high fidelity gate operations between distant quantum nodes and the creation of multi-photon entangled states. Rather than solely being an effect which must be corrected for, dispersion also opens up new possibilities to engineer light-matter interactions with exotic properties~\cite{sundaresanInteractingQubitPhotonBound2019,kimQuantumElectrodynamicsTopological2021a,gonzalez-tudelaSubwavelengthVacuumLattices2015, lodahlControllingDynamicsSpontaneous2004}. Additionally, dispersion enables the propagation of self-focusing pulses,~\cite{boydSelffocusingPresentFundamentals2009}, demonstrated in pulse compression schemes~\cite{brabecIntenseFewcycleLaser2000a} that have revolutionized the field of attosecond physics~\cite{krauszAttosecondPhysics2009}. Focusing pulses through dispersion offers exciting and novel possibilities to control quantum systems, which have not been experimentally explored. Recently, it was proposed to use a self-compressing pulse to address individual quantum emitters~\cite{casullerasRemoteIndividualAddressing2021}.

\begin{figure*}[ht!]
    \includegraphics[width=\linewidth]{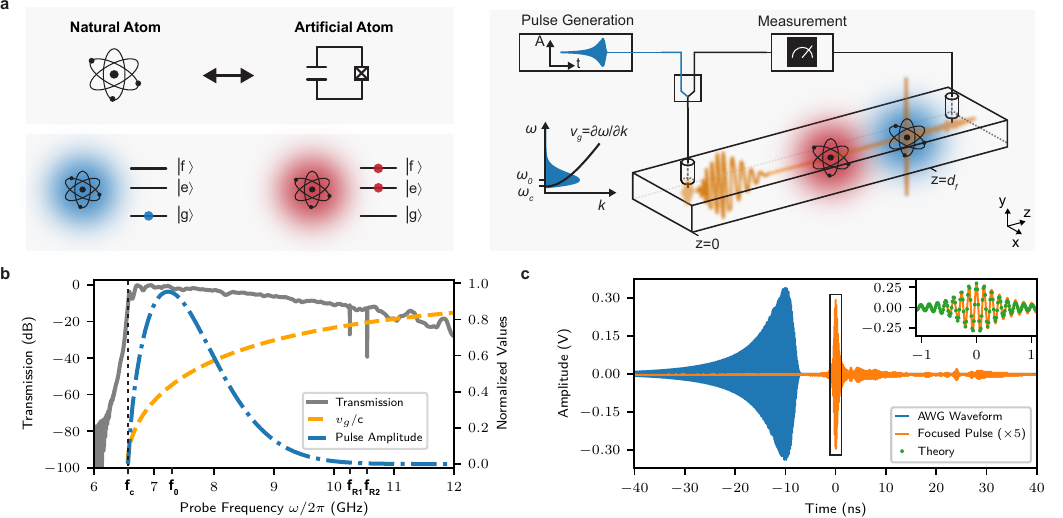}
    \caption{
    \textbf{Self-focusing pulses to control artificial atoms.}
    \textbf{a} Superconducting qubits (e.g. transmons) are artificial atomic systems whose quantum states may be coherently controlled by microwave pulses. In this experiment, two transmon qubits (Q1 and Q2), coupled to their respective readout resonators, are placed at separate locations along a rectangular waveguide. The input of the waveguide is connected to an arbitrary waveform generator which creates a chirped microwave pulse (blue), whose envelope is illustrated in both the time and frequency domains. The self-compressing pulse is illustrated in space (orange). Depending on the location of the qubits in the waveguide, this pulse either keeps the qubit in its ground state $\ket{g}$ (blue atom) or excites it to higher states $\ket{e}$ and $\ket{f}$ (red atom).
    \textbf{b} Transmission $|S_{21}|$ (grey, left axis) through a microwave waveguide, measured at the \SI{20}{\milli\kelvin} stage of a cryostat. The cutoff frequency is indicated by $f_{\rm c}$ (dashed black line) and the readout resonator frequencies of Q1 and Q2 by $f_{R1}$ and $f_{R2}$, respectively. Group velocity of the waveguide (orange), normalized by the vacuum speed of light, $c$ (right axis). The spectrum of the wideband, normalized pulse amplitude (blue) is shown for a central frequency of $\omega_{0}/2\pi=\SI{7.2}{\giga\hertz}$ and a spot size of $\sigma_f=\SI{3.5}{\centi\meter}$ (right axis). 
    \textbf{c} Room-temperature demonstration of pulse focusing. The generated pulse with the displayed waveform (blue) is injected at the input of a rectangular waveguide. After propagation through the waveguide, the self-focused pulse is measured at the focal point with a weakly coupled antenna (orange).
    Inset: Comparison of the theoretically expected (green dots) and the measured (orange line) pulse amplitude .
    }
    \label{fig:Fig1}
\end{figure*}

Here, we use dispersion to create fast, localized microwave pulses to selectively control superconducting qubits within a waveguide. Specifically, we demonstrate this technique using the non-linear dispersion around the cutoff frequency of a microwave waveguide, where the group velocity has a strong spectral dependence. By shaping the phase and amplitude of a chirped microwave pulse within a bandwidth of several GHz, we induce a self-focusing effect caused by the dispersion. This self-focusing phenomenon allows us to achieve individual addressing of two superconducting transmon qubits, denoted as Q1 and Q2, separated  by a sub-wavelength distance. This experiment demonstrates a novel control scheme for qubits embedded in dispersive media, which is especially useful for systems in which local qubit control is difficult to engineer.

Specifically, we consider a rectangular waveguide, where only transverse electric (TE) waves and transverse magnetic (TM) waves propagate through the waveguide vacuum~\cite{pozarMicrowaveEngineering2012}. The cutoff frequencies of the propagating modes and their non-linear dispersion relation are governed by the geometry of the waveguide boundaries. The dispersion relation of the waveguide for a propagating mode with wavevector $k=2\pi/\lambda$, where $\lambda$ is the wavelength inside the waveguide, is given by
\begin{equation}
    \omega(k) = c\sqrt{k^2+k_c^2}.
    \label{eq:dispersion}
\end{equation}
Here, $\omega$ is the angular frequency, $k_c= \omega_c/c$ is the cutoff wavevector ($\omega_c = 2\pi f_{\rm c}$ is the cutoff angular frequency), and $c$ is the speed of light in the material within the waveguide (in this case vacuum). In~\autoref{fig:Fig1}b, we plot the analytical frequency dependence of the group velocity for the first mode TE$_{10}$ of a WR90 rectangular waveguide~\cite{pozarMicrowaveEngineering2012}. This mode has a cutoff frequency given by $\omega_{\rm c}/2\pi =\SI{6.56}{\giga \hertz}$, which is visible when measuring the transmission through the waveguide, shown in~\autoref{fig:Fig1}b. For frequencies above the cutoff, the group velocity increases following~\autoref{eq:dispersion}, and in the high-frequency limit asymptotes to $c$. Thus, the individual spectral components of a chirped pulse propagate with different velocities. In that case, the pulse parameters can be adjusted to produce a strong focusing effect at a specific focal point (see \autoref{fig:Fig1}a). In particular, the electrical field amplitude of a self-compressing pulse at the focal point $d_f$ can be written as~\cite{casullerasRemoteIndividualAddressing2021}
\begin{equation}
     E_f(t) = E_0 \int_{0}^{\infty} \text{d}k~\tilde{E}(k) \cos \left(\omega(k) t\right),
     \label{eq:pulse_definition}
\end{equation}
\begin{equation*}
        \textrm{with} \quad \tilde{E}(k) = \exp \left( -\frac{\sigma_f^2(k-k_0)^2}{2} \right),
\end{equation*}
where $E_0$ is the pulse amplitude, $\sigma_f$ is the pulse spot size, and $k_0$ is the central wavevector resulting in the central angular frequency $\omega_{0}=\omega(k_0)$. The amplitude of the pulse in the spectral domain is shown in~\autoref{fig:Fig1}b. Note that we can compute the pulse $E(t,z,d_f)$ at any position $z$ along the waveguide and choose an arbitrary focal point $d_f$ by changing the phase of the pulse $E_f(t)$. By convention, the input of the waveguide is set to $z=0$ (see Supplementary Material for details).

We first demonstrate the self-focusing effect of a chirped pulse in a room temperature experiment where the wave packet, generated by a fast-sampling arbitrary waveform generator (AWG), propagates \SI{1.03}{\meter} through a rectangular waveguide. The pulse parameters are chosen such that the focal point is at a weakly coupled antenna, which is connected to an oscilloscope. In~\autoref{fig:Fig1}b, we display the waveform injected at the input of the waveguide for a chirped pulse with a central frequency of $\omega_{0}/2\pi=\SI{7.2}{\giga\hertz}$ and spot size of $\sigma_f= \SI{3.5}{\centi\meter}$. The central frequency is chosen to maximize the self-focusing effect while minimizing the effect of the strong reflections from the waveguide couplers near the cutoff frequency. In~\autoref{fig:Fig1}c, we compare the temporal full width at half maximum (FWHM) of the pulse at the output of the generator $\sigma_{\mathrm{AWG}} = \SI{10}{\nano\second}$ with the pulse at the focal point $\sigma_{fp} < \SI{1}{\nano\second}$, showing a pulse compression by more than a factor of 10. The measurement at the antenna position is then compared to the expected waveform given by~\autoref{eq:pulse_definition}. The experimental results show remarkable agreement with the model, considering that only the time delay due to the connecting cable and the pulse amplitude due to losses in the experiment are adjusted. This experiment conclusively demonstrates the self-focusing effect in the time domain.

To understand the ability to selectively addressing a quantum emitter using a time-focused pulse, we first describe the dynamics of a qubit interacting with an ideal self-focusing pulse. When a qubit, initially in the ground state, is positioned at the focal point of the pulse, it is excited and subsequently de-excited as the pulse passes~\cite{casullerasRemoteIndividualAddressing2021}. Conversely, a qubit located out of focus transitions to the excited state and remains excited after the pulse interaction. This behavior arises from the interplay between the pulse's instantaneous amplitude and the qubit's detuning, modeled as a Landau-Zener process with a time-dependent detuning and Rabi coupling strength (see Supplementary Material for more details). The qubit eigenenergies exhibit an avoided crossing, with an energy gap proportional to the time-dependent electric field strength, which causes the dynamics described above.

During the generation of the chirped self-compressing pulse, the timing can be parameterized such that the focal point coincides with the position of a single qubit, while for all other locations (e.g. the position of a second qubit) the pulse is out of focus  (see~\autoref{fig:Fig1}a). In that case, the target qubit remains in its ground state as long as it lies within the spatial resolution of the pulse, denoted by $\sigma_q$, while the other qubits remain excited. Here, we define $\sigma_q$ as the FWHM of the probability of the qubit to be in the ground state as the focal point $d_f$ of the pulse is swept.


To experimentally investigate the interaction of the self-focusing pulse with a quantum emitter and to explore the possibility to individually address spatially separated qubits, we design the waveguide QED setup depicted in~\autoref{fig:Fig1}a. Two superconducting transmon qubits are embedded in a rectangular waveguide~\cite{kochChargeinsensitiveQubitDesign2007b,pozarMicrowaveEngineering2012}. The transmon qubits act as multi-level artificial atoms while the waveguide provides a dispersive medium. The spatial separation between the qubits along the pulse propagation direction corresponds to half a wavelength, $d_z = \SI{5}{\centi\meter} = \lambda/2$, at $\SI{7.2}{\giga\hertz}$. Each qubit is additionally coupled to a readout-resonator to individually determine the qubit state~\cite{zoepflCharacterizationLowLoss2017}. The fundamental resonance frequencies of the readout resonators, $\omega_{R1}/2\pi = \SI{10.25}{\giga \hertz}$ and $\omega_{R2}/2\pi = \SI{10.55}{\giga \hertz}$, are visible in the transmission measurement in~\autoref{fig:Fig1}b. The qubits can be tuned such that their transition frequency $\omega_{ge}/2\pi$ matches the central frequency of the microwave pulse $\omega_{0}/2\pi$. 

\begin{figure*}[ht]
    \centering
    \includegraphics[width=1\linewidth]{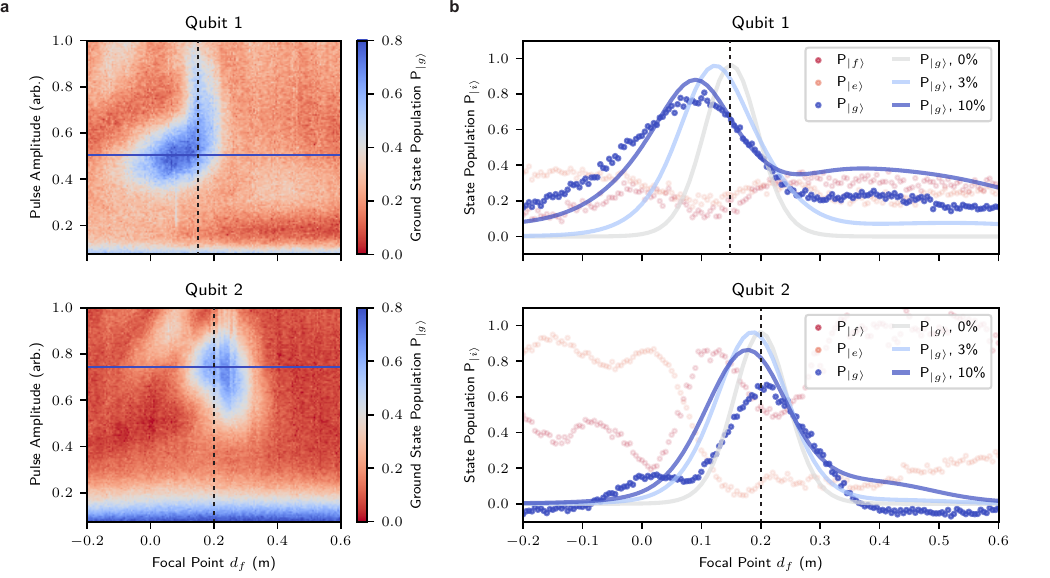}
    \caption{\textbf{Spatial addressing of a single qubit.}
   \textbf{a} Probability of measuring Q1 or Q2 in the ground state after the interaction with a wideband pulse, as a function of the focal point $d_f$ and the spot size $\sigma_f$. For each panel, the frequency of Q1 or Q2 is tuned to $\omega_e/2\pi=\omega_0/2\pi$, while the other qubit is tuned below the waveguide cutoff to limit its interaction with the pulse. \textbf{b} Linecuts of the ground state population $P_{|g\rangle}$ (blue solid lines in \textbf{a}) for Q1 and Q2, respectively, for the optimal pulse amplitude. The plots show the measured (data points) and simulated (solid lines) population $P_{|g\rangle}$ as a function of the focal point. The simulations show the distortion of the lineshape when changing from an ideal waveguide to realistic estimates of the reflected power (in \%) at the output. The qubit locations, determined with a tape measure at room temperature, are indicated by the dashed vertical lines. In addition, we show the measured first $P_{|e\rangle}$ and second $P_{|f\rangle}$ excited state populations for each qubit. Note that the state populations sum to more than 1 due to calibration instabilities (see Supplementary Material for details). 
    }
    \label{fig:Fig2}
\end{figure*}

To demonstrate spatial addressing we first show that the population of a single qubit can be controlled by its interaction with the pulse. Subsequently, we extend this demonstration to show spatial addressing for a two-qubit system. To this end, we first tune the frequency of a single qubit inside the waveguide band ($\omega_{ge}/2\pi = \omega_{0}/2\pi=\SI{7.2}{\giga \hertz}$) while the second qubit is tuned below the cutoff frequency. Hence, the second qubit does not interact with propagating electromagnetic waves. We send a frequency chirped pulse with fixed spot size $\sigma_f = \SI{3.5}{cm}$ and vary its focal point and amplitude. In~\autoref{fig:Fig2}a, we show the resulting ground state population of both qubits when interacting individually with the addressing pulse. For small pulse amplitudes both qubits remain mainly in the ground state (blue region close to the x-axis). When increasing the pulse amplitude, we observe that most of the qubit population remains in the ground state only if the focal point of the pulse coincides with the position of the qubit. However, for the other focal points, the ground state population decreases due to the excitation of higher states of the transmon qubit. There exists an optimal pulse amplitude where this behaviour occurs, as theoretically predicted~\cite{casullerasRemoteIndividualAddressing2021}. The optimal amplitude for Q1 is lower than for Q2 as it is coupled stronger to the waveguide. In particular, for the optimal pulse amplitude, the qubit remains in the ground state for a focal points of $d_{f} \approx \SI{0.1}{\meter}$ for Q1, and at $d_{f}\approx \SI{0.2}{\meter}$ for Q2, corresponding to the positions of the qubit in the waveguide. 

To further illustrate this behavior,~\autoref{fig:Fig2}b shows the population of the ground state $P_{|g\rangle}$, as well as the first $P_{|e\rangle}$ and second excited state $P_{|f\rangle}$ for both qubits, as a function of the focal point for the optimal pulse amplitude. 
These populations slightly deviate from those corresponding to the interaction with an ideal pulse~\cite{casullerasRemoteIndividualAddressing2021}. In particular, both data and simulation of the ground state show a shift of the center of the probability distribution with respect to the qubit positions, as well as a decrease of the maximum ground state population. This discrepancy is due to experimental imperfections. Specifically, the broadband pulse interacts with the microwave wiring and filtering in the cryostat, whose frequency response distorts the pulse shape before it arrives at the waveguide input. 
In addition, after propagating through the waveguide, the pulse is partially reflected at the waveguide output due to imperfect impedance matching. The reflected signal  is evident from the reoccurring delayed amplitude peaks for $\rm t > 0$ in~\autoref{fig:Fig1}c (orange trace). The distortion and reflection of the pulse ultimately result in unwanted qubit dynamics, limiting the state preparation fidelity, spatial resolution and accuracy of the focal point in~\autoref{fig:Fig2}b. To model the effect of the reflections at the waveguide output, we simulate the qubit dynamics using an ideal pulse subject to a point-like reflection (see~\autoref{fig:Fig2}b). The reflection points are located at $\SI{10}{cm}$ and $\SI{5}{cm}$ after Q1 and Q2, respectively, which corresponds to the end of the waveguide in the experimental setup. For the focal points indicated by the vertical dashed lines in~\autoref{fig:Fig2}b, we compare the experimental data to models with a broadband reflected power of 3\% and 10\% at the waveguide output (corresponding to a return loss of respectively \SI{-15}{\decibel} and \SI{-10}{\decibel}). For greater reflected power, we observe a stronger distortion of the lineshape, as expected, emphasizing the need for better impedance-matching of the waveguide couplers. As shown in~\autoref{fig:Fig2}b, the ground state population of Q1 qualitatively agrees with the simulations for a reflected power of 10\%. Additionally, note that we observe a stronger impact of the reflections on Q1 than on Q2 due to the longer round-trip distance that a reflected pulse must travel before returning to Q1 (see Supplementary Material for further discussion).

\begin{figure*}[ht!]
    \centering\includegraphics[width=1\linewidth]{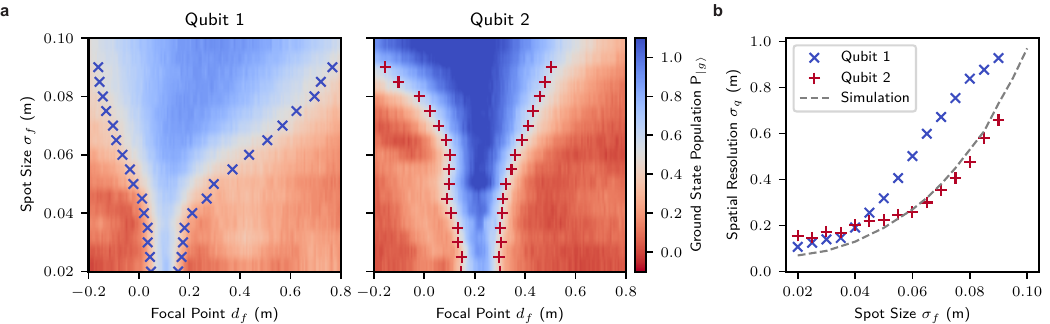}
    \caption{
    \textbf{Spatial resolution of the qubit addressing.}
    \textbf{a} Ground state population of Q1 and Q2 as a function of the spot size $\sigma_f$ and the focal point $d_f$ of the pulse. For each panel, only the frequency of the corresponding qubit is tuned inside the waveguide band. The red and blue crosses indicate the full-width at half maximum (FWHM) of the spatial resolution $\sigma_q$, extracted from the experimental data. As the spot size decreases from $\sigma_f = \SI{0.1}{\meter}$ to $\sigma_f=\SI{0.02}{\meter}$, the spatial resolution $\sigma_q$ is reduced, reaching a minimum value of $\sigma_q \approx \SI{15}{cm}$. \textbf{b} Measured spatial resolution $\sigma_q$ for Q1 and Q2 as a function of to the spot size $\sigma_f$ of the pulse, compared to the simulation of an ideal pulse interacting with the qubit (grey). The extracted FWHM of the experiment can deviate from the simulation due to imperfections of the waveguide input and output (see~\autoref{fig:Fig2}).
    }
    \label{fig:Fig3}
\end{figure*}

We demonstrate control over the spatial resolution $\sigma_q$ of the local addressing by varying the spot size $\sigma_f$ of the focusing pulse. In particular, ~\autoref{fig:Fig3}a shows the ground state population of Q1 and Q2 as a function of $\sigma_f$ for the optimal pulse amplitude. For large values of $\sigma_f$, we see observe a weak dependence of $\sigma_q$ on the focal point, similar to a conventional Gaussian pulse excitation, which cannot realize a position dependence. When decreasing $\sigma_f$, the population in the ground state shows a stronger dependence on the focal point, effectively reducing $\sigma_q$. To experimentally verify the scaling of $\sigma_q$, we extract the FWHM of the measured ground state population as a function of the spot size, shown in~\autoref{fig:Fig3}b. For small values of $\sigma_f$, the spatial resolution saturates to $\sigma_q \approx \SI{15}{\centi\meter}$, which is limited by the finite speed of light, $c$, and by the pulse distortion and reflections. Note that for the experiments shown in~\autoref{fig:Fig2}, we choose $\sigma_f = \SI{3.5}{\centi\meter}$ as the best compromise between a good spatial resolution and a maximum population of the ground state at the focal point. The simulation uses an ideal pulse interacting with a qubit at the focal point and emphasizes the susceptibility to experimental imperfections.

\begin{figure}[ht!]
    \centering
    \includegraphics[width=\linewidth]{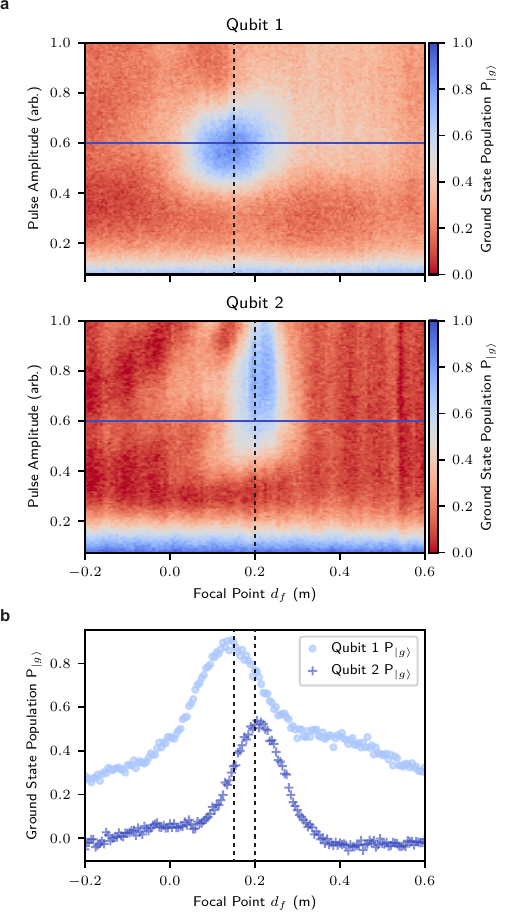}
    \caption{
    \textbf{Spatial addressing of two resonant qubits.}
    \textbf{a} Ground state population of Q1 and Q2 after interaction with a self-focusing pulse, as a function of the focal point $d_f$ and spot size $\sigma_f$ of the pulse. Here, the frequencies of both qubits are tuned to $\omega_e/2\pi=\omega_0/2\pi$, where $\omega_0$ is the central frequency of the pulse. The dashed vertical lines indicate the qubit locations. \textbf{b} Linecuts (blue solid lines in \textbf{a}) showing the ground state population for Q1 and Q2, for the optimal pulse amplitude associated to Q1. These measurements indicate the ability to selectively address the individual qubits with the pulse parameter $d_f$. To improve the addressability it is possible to use different pulse amplitudes, optimized for Q1 and Q2.}
    \label{fig:Fig4}
\end{figure}

To demonstrate spatial addressing for a two-qubit system, we tune both qubits to $\omega_{\rm ge}/2\pi = \SI{7.28}{\giga \hertz}$. As both qubits are coupled to the waveguide, photon-mediated interactions form a hybridized qubit system~\cite{lalumiereInputoutputTheoryWaveguide2013b,zannerCoherentControlMultiqubit2022a}. However, here the timescale of this interaction, governed by the coupling strength between the qubit and the waveguide $\Gamma_{q}/2\pi \approx \SI{30}{\kilo \Hz}$, is very long compared to the length of the excitation pulse, $t= \SI{100}{\nano\second} \ll 1/\Gamma_q$, and thus we disregard formations of collective states of the qubit pair during the experiment. \autoref{fig:Fig4}a shows the ground state population of Q1 and Q2 following the experiment in~\autoref{fig:Fig2}, now for both qubits tuned into resonance. Even in this scenario, the self-focusing pulse spatially addresses both qubits separately. As mentioned previously, distortions of the pulse can lead to significant changes in the qubit response and shift the focal point (see Supplementary Material). Thus, we attribute the different behavior of the two-qubit system compared to the single-qubit case to different transmission characteristics at \SI{7.28}{GHz}. To further illustrate the ability to selectively address the qubits individually, we show the evolution of the ground state population $P_{\ket{g}}$ of Q1 and Q2 as a function of the focal point, for the optimal pulse amplitude associated to Q1 (\autoref{fig:Fig4}b). 
The population for each qubit is similar to the results of the single qubit experiment, shown in~\autoref{fig:Fig2}. As before,  $P_{\ket{g}}$ for Q1 reaches a maximum for a focal point of $d_f \approx \SI{0.15}{\meter}$, while $P_{\ket{g}}$ for Q2 is maximal at $d_f \approx \SI{0.2}{\meter}$, corresponding to the qubit locations. We remark that the state preparation fidelity for Q2 can be improved by using a drive amplitude optimized for this qubit. Also note that each of the qubits can be prepared in either the ground state or an excited state, by adjusting the focal point $d_f$ and amplitude of the pulse. Given the physical separation of \SI{5}{cm} between the qubits, which corresponds to approximately half of a wavelength in the waveguide at the qubit frequency, the measurements presented in~\autoref{fig:Fig4} demonstrate the ability to individually address qubits within a sub-wavelength distance.

Our experimental findings represent the first realization of position-dependent qubit addressing using self-compressing pulses. We achieve focusing by generating a chirped broad-band pulse that propagates through a dispersive medium, namely a rectangular microwave waveguide. Manipulating the phases of the pulse allows to choose the focal point, enabling selective excitation of two qubits separated by a distance smaller than the wavelength. We demonstrate spatial addressing both in the single-qubit scenario, where only one qubit interacts with the pulse, and in the two-qubit system in a weakly hybridized regime, where both qubits are resonant with the central frequency of the pulse. Furthermore, we achieve a minimal spatial resolution of $\sigma_q \approx \SI{15}{cm}$, limited by the dispersion of the waveguide with a finite speed of light and experimental imperfections. Analogous to classical high-speed communication, our results emphasize the crucial role of dispersion for quantum physics experiments when using high-speed broad-band pulses and their potential applications.

 The focusing technique, based on tailoring the dispersion of microwave components, can be employed in quantum communication and quantum simulation experiments where multiple qubits are coupled to a waveguide and local control is not directly engineerable. Locally exciting a chain of qubits without the need to physically adapt the experimental architecture provides a powerful tool to allow state preparation in Ising-physics simulations~\cite{brehmWaveguideBandgapEngineering2021,poshakinskiyQuantumChaosDriven2021} or to drive multi-qubit dark states~\cite{zannerCoherentControlMultiqubit2022a}. We expect that simple modifications such as optimizing the impedance-matching of the waveguide-coaxial adapters around the qubit frequencies, improving the qubit coherence by operating at a flux sweet-spot and enhancing the qubit-waveguide coupling will improve the preparation fidelity and the spatial resolution. In addition, the current experiment shows limitations caused by the small anharmonicity of the transmon qubits used, which leads to the driving of multi-photon transitions and leakage out of the qubit states. This problem can be avoided by utilizing other types of qubits, e.g. a fluxonium~\cite{manucharyanFluxoniumSingleCooperPair2009b} which typically has a greater anharmonicity. 
 
 Furthermore, this technique can  be extended to other types of atomic systems, including solid-state spin qubits, such as nitrogen-vacancy centers in diamond~\cite{wolfowiczQuantumGuidelinesSolidstate2021}. In fact, solid-state spin qubits could benefit greatly from the self-focusing approach because their point-dipole nature makes spatially close qubits difficult to individually address with microwave pulses. Additionally, solid-state spin qubits can have a high anharmonicity compared to superconducting qubits, which would reduce the population in higher-order levels and thus allow realization of this technique with high precision. In conclusion, the sub-wavelength control developed in our work provides an avenue to achieve precise manipulation of individual qubits within densely packed registers. The technique of sub-wavelength focusing holds broad applications across different quantum platforms and frequency domains, enhancing overall qubit control capabilities.

\fakesection{Additional Information}
\textbf{Data availability}
The data that support the findings of this study are available on Zenodo.
\newline
\textbf{Code availability}
The code used for the data analysis and simulated results is available on Zenodo.
\newline
\textbf{Acknowledgments}
We thank Carlos Gonzalez-Ballestero for valuable input and discussions about the self-focusing theory. We thank Andreas Strasser for fabricating the waveguide. M.Z. acknowledges funding by the European Research Council (ERC) under the European Unions Horizon 2020 research and innovation program (714235). This research was funded in part by the Austrian Science Fund (FWF) DOI 10.55776/W1259 as well as DOI 10.55776/F71. For the purpose of open access, the author has applied a CC BY public copyright licence to any Author Accepted Manuscript version arising from this submission. E. I. R. acknowledges support by an appointment to the Intelligence Community Postdoctoral Research Fellowship Program at Stanford University administered by Oak Ridge Institute for Science and Education (ORISE) through an interagency agreement between the U.S. Department of Energy and the Office of the Director of National Intelligence (ODNI).
\newline
\textbf{Competing interests:} The authors declare no competing interests.
\newline
\textbf{Author  contributions:} M.Z., R.A. and G.K. conceived and designed the experiment. M.Z. simulated and M.Z., C.M.F.S., I.Y. fabricated the devices. M.Z., R.A. and E.I.R. conducted the measurements. M.Z. and R.A. analyzed the data. S.C., R.A. and  O. R.-I. developed the theoretical model. M.Z., R.A., E.I.R. and G.K. wrote the manuscript. All authors discussed the results and contributed to the writing of the manuscript.
\newline
\textbf{Correspondence and requests for materials} should be addressed to G.K.
\bibliography{references}

\begin{thebibliography}{32}%
\makeatletter
\providecommand \@ifxundefined [1]{%
 \@ifx{#1\undefined}
}%
\providecommand \@ifnum [1]{%
 \ifnum #1\expandafter \@firstoftwo
 \else \expandafter \@secondoftwo
 \fi
}%
\providecommand \@ifx [1]{%
 \ifx #1\expandafter \@firstoftwo
 \else \expandafter \@secondoftwo
 \fi
}%
\providecommand \natexlab [1]{#1}%
\providecommand \enquote  [1]{``#1''}%
\providecommand \bibnamefont  [1]{#1}%
\providecommand \bibfnamefont [1]{#1}%
\providecommand \citenamefont [1]{#1}%
\providecommand \href@noop [0]{\@secondoftwo}%
\providecommand \href [0]{\begingroup \@sanitize@url \@href}%
\providecommand \@href[1]{\@@startlink{#1}\@@href}%
\providecommand \@@href[1]{\endgroup#1\@@endlink}%
\providecommand \@sanitize@url [0]{\catcode `\\12\catcode `\$12\catcode `\&12\catcode `\#12\catcode `\^12\catcode `\_12\catcode `\%12\relax}%
\providecommand \@@startlink[1]{}%
\providecommand \@@endlink[0]{}%
\providecommand \url  [0]{\begingroup\@sanitize@url \@url }%
\providecommand \@url [1]{\endgroup\@href {#1}{\urlprefix }}%
\providecommand \urlprefix  [0]{URL }%
\providecommand \Eprint [0]{\href }%
\providecommand \doibase [0]{https://doi.org/}%
\providecommand \selectlanguage [0]{\@gobble}%
\providecommand \bibinfo  [0]{\@secondoftwo}%
\providecommand \bibfield  [0]{\@secondoftwo}%
\providecommand \translation [1]{[#1]}%
\providecommand \BibitemOpen [0]{}%
\providecommand \bibitemStop [0]{}%
\providecommand \bibitemNoStop [0]{.\EOS\space}%
\providecommand \EOS [0]{\spacefactor3000\relax}%
\providecommand \BibitemShut  [1]{\csname bibitem#1\endcsname}%
\let\auto@bib@innerbib\@empty
\bibitem [{\citenamefont {Brabec}\ and\ \citenamefont {Krausz}(2000)}]{brabecIntenseFewcycleLaser2000a}%
  \BibitemOpen
  \bibfield  {author} {\bibinfo {author} {\bibfnamefont {T.}~\bibnamefont {Brabec}}\ and\ \bibinfo {author} {\bibfnamefont {F.}~\bibnamefont {Krausz}},\ }\bibfield  {title} {\bibinfo {title} {Intense few-cycle laser fields: {{Frontiers}} of nonlinear optics},\ }\href {https://doi.org/10.1103/RevModPhys.72.545} {\bibfield  {journal} {\bibinfo  {journal} {Reviews of Modern Physics}\ }\textbf {\bibinfo {volume} {72}},\ \bibinfo {pages} {545} (\bibinfo {year} {2000})}\BibitemShut {NoStop}%
\bibitem [{\citenamefont {Pozar}(2012)}]{pozarMicrowaveEngineering2012}%
  \BibitemOpen
  \bibfield  {author} {\bibinfo {author} {\bibfnamefont {D.~M.}\ \bibnamefont {Pozar}},\ }\href@noop {} {\emph {\bibinfo {title} {Microwave Engineering}}},\ \bibinfo {edition} {4th}\ ed.\ (\bibinfo  {publisher} {Wiley},\ \bibinfo {address} {Hoboken, NJ},\ \bibinfo {year} {2012})\BibitemShut {NoStop}%
\bibitem [{\citenamefont {Joannopoulos}\ \emph {et~al.}(1997)\citenamefont {Joannopoulos}, \citenamefont {Villeneuve},\ and\ \citenamefont {Fan}}]{joannopoulosPhotonicCrystalsPutting1997c}%
  \BibitemOpen
  \bibfield  {author} {\bibinfo {author} {\bibfnamefont {J.~D.}\ \bibnamefont {Joannopoulos}}, \bibinfo {author} {\bibfnamefont {P.~R.}\ \bibnamefont {Villeneuve}},\ and\ \bibinfo {author} {\bibfnamefont {S.}~\bibnamefont {Fan}},\ }\bibfield  {title} {\bibinfo {title} {Photonic crystals: Putting a new twist on light},\ }\href {https://doi.org/10.1038/386143a0} {\bibfield  {journal} {\bibinfo  {journal} {Nature}\ }\textbf {\bibinfo {volume} {386}},\ \bibinfo {pages} {143} (\bibinfo {year} {1997})}\BibitemShut {NoStop}%
\bibitem [{\citenamefont {Sheremet}\ \emph {et~al.}(2023)\citenamefont {Sheremet}, \citenamefont {Petrov}, \citenamefont {Iorsh}, \citenamefont {Poshakinskiy},\ and\ \citenamefont {Poddubny}}]{sheremetWaveguideQuantumElectrodynamics2023}%
  \BibitemOpen
  \bibfield  {author} {\bibinfo {author} {\bibfnamefont {A.~S.}\ \bibnamefont {Sheremet}}, \bibinfo {author} {\bibfnamefont {M.~I.}\ \bibnamefont {Petrov}}, \bibinfo {author} {\bibfnamefont {I.~V.}\ \bibnamefont {Iorsh}}, \bibinfo {author} {\bibfnamefont {A.~V.}\ \bibnamefont {Poshakinskiy}},\ and\ \bibinfo {author} {\bibfnamefont {A.~N.}\ \bibnamefont {Poddubny}},\ }\bibfield  {title} {\bibinfo {title} {Waveguide quantum electrodynamics: {{Collective}} radiance and photon-photon correlations},\ }\href {https://doi.org/10.1103/RevModPhys.95.015002} {\bibfield  {journal} {\bibinfo  {journal} {Reviews of Modern Physics}\ }\textbf {\bibinfo {volume} {95}},\ \bibinfo {pages} {015002} (\bibinfo {year} {2023})}\BibitemShut {NoStop}%
\bibitem [{\citenamefont {Masson}\ and\ \citenamefont {{Asenjo-Garcia}}(2020)}]{massonAtomicwaveguideQuantumElectrodynamics2020}%
  \BibitemOpen
  \bibfield  {author} {\bibinfo {author} {\bibfnamefont {S.~J.}\ \bibnamefont {Masson}}\ and\ \bibinfo {author} {\bibfnamefont {A.}~\bibnamefont {{Asenjo-Garcia}}},\ }\bibfield  {title} {\bibinfo {title} {Atomic-waveguide quantum electrodynamics},\ }\href {https://doi.org/10/gh2n43} {\bibfield  {journal} {\bibinfo  {journal} {Physical Review Research}\ }\textbf {\bibinfo {volume} {2}},\ \bibinfo {pages} {043213} (\bibinfo {year} {2020})}\BibitemShut {NoStop}%
\bibitem [{\citenamefont {Astafiev}\ \emph {et~al.}(2010)\citenamefont {Astafiev}, \citenamefont {Zagoskin}, \citenamefont {Abdumalikov}, \citenamefont {Pashkin}, \citenamefont {Yamamoto}, \citenamefont {Inomata}, \citenamefont {Nakamura},\ and\ \citenamefont {Tsai}}]{astafievResonanceFluorescenceSingle2010b}%
  \BibitemOpen
  \bibfield  {author} {\bibinfo {author} {\bibfnamefont {O.}~\bibnamefont {Astafiev}}, \bibinfo {author} {\bibfnamefont {A.~M.}\ \bibnamefont {Zagoskin}}, \bibinfo {author} {\bibfnamefont {A.~A.}\ \bibnamefont {Abdumalikov}}, \bibinfo {author} {\bibfnamefont {Y.~A.}\ \bibnamefont {Pashkin}}, \bibinfo {author} {\bibfnamefont {T.}~\bibnamefont {Yamamoto}}, \bibinfo {author} {\bibfnamefont {K.}~\bibnamefont {Inomata}}, \bibinfo {author} {\bibfnamefont {Y.}~\bibnamefont {Nakamura}},\ and\ \bibinfo {author} {\bibfnamefont {J.~S.}\ \bibnamefont {Tsai}},\ }\bibfield  {title} {\bibinfo {title} {Resonance {{Fluorescence}} of a {{Single Artificial Atom}}},\ }\href {https://doi.org/10/bd5fq5} {\bibfield  {journal} {\bibinfo  {journal} {Science}\ }\textbf {\bibinfo {volume} {327}},\ \bibinfo {pages} {840} (\bibinfo {year} {2010})}\BibitemShut {NoStop}%
\bibitem [{\citenamefont {van Loo}\ \emph {et~al.}(2013)\citenamefont {van Loo}, \citenamefont {Fedorov}, \citenamefont {Lalumi{\`e}re}, \citenamefont {Sanders}, \citenamefont {Blais},\ and\ \citenamefont {Wallraff}}]{looPhotonMediatedInteractionsDistant2013a}%
  \BibitemOpen
  \bibfield  {author} {\bibinfo {author} {\bibfnamefont {A.~F.}\ \bibnamefont {van Loo}}, \bibinfo {author} {\bibfnamefont {A.}~\bibnamefont {Fedorov}}, \bibinfo {author} {\bibfnamefont {K.}~\bibnamefont {Lalumi{\`e}re}}, \bibinfo {author} {\bibfnamefont {B.~C.}\ \bibnamefont {Sanders}}, \bibinfo {author} {\bibfnamefont {A.}~\bibnamefont {Blais}},\ and\ \bibinfo {author} {\bibfnamefont {A.}~\bibnamefont {Wallraff}},\ }\bibfield  {title} {\bibinfo {title} {Photon-{{Mediated Interactions Between Distant Artificial Atoms}}},\ }\href {https://doi.org/10.1126/science.1244324} {\bibfield  {journal} {\bibinfo  {journal} {Science}\ ,\ \bibinfo {pages} {1244324}} (\bibinfo {year} {2013})}\BibitemShut {NoStop}%
\bibitem [{\citenamefont {Zanner}\ \emph {et~al.}(2022)\citenamefont {Zanner}, \citenamefont {Orell}, \citenamefont {Schneider}, \citenamefont {Albert}, \citenamefont {Oleschko}, \citenamefont {Juan}, \citenamefont {Silveri},\ and\ \citenamefont {Kirchmair}}]{zannerCoherentControlMultiqubit2022a}%
  \BibitemOpen
  \bibfield  {author} {\bibinfo {author} {\bibfnamefont {M.}~\bibnamefont {Zanner}}, \bibinfo {author} {\bibfnamefont {T.}~\bibnamefont {Orell}}, \bibinfo {author} {\bibfnamefont {C.~M.~F.}\ \bibnamefont {Schneider}}, \bibinfo {author} {\bibfnamefont {R.}~\bibnamefont {Albert}}, \bibinfo {author} {\bibfnamefont {S.}~\bibnamefont {Oleschko}}, \bibinfo {author} {\bibfnamefont {M.~L.}\ \bibnamefont {Juan}}, \bibinfo {author} {\bibfnamefont {M.}~\bibnamefont {Silveri}},\ and\ \bibinfo {author} {\bibfnamefont {G.}~\bibnamefont {Kirchmair}},\ }\bibfield  {title} {\bibinfo {title} {Coherent control of a multi-qubit dark state in waveguide quantum electrodynamics},\ }\href {https://doi.org/10.1038/s41567-022-01527-w} {\bibfield  {journal} {\bibinfo  {journal} {Nature Physics}\ }\textbf {\bibinfo {volume} {18}},\ \bibinfo {pages} {538} (\bibinfo {year} {2022})}\BibitemShut {NoStop}%
\bibitem [{\citenamefont {Hoi}\ \emph {et~al.}(2011)\citenamefont {Hoi}, \citenamefont {Wilson}, \citenamefont {Johansson}, \citenamefont {Palomaki}, \citenamefont {Peropadre},\ and\ \citenamefont {Delsing}}]{hoiDemonstrationSinglePhotonRouter2011}%
  \BibitemOpen
  \bibfield  {author} {\bibinfo {author} {\bibfnamefont {I.-C.}\ \bibnamefont {Hoi}}, \bibinfo {author} {\bibfnamefont {C.~M.}\ \bibnamefont {Wilson}}, \bibinfo {author} {\bibfnamefont {G.}~\bibnamefont {Johansson}}, \bibinfo {author} {\bibfnamefont {T.}~\bibnamefont {Palomaki}}, \bibinfo {author} {\bibfnamefont {B.}~\bibnamefont {Peropadre}},\ and\ \bibinfo {author} {\bibfnamefont {P.}~\bibnamefont {Delsing}},\ }\bibfield  {title} {\bibinfo {title} {Demonstration of a {{Single-Photon Router}} in the {{Microwave Regime}}},\ }\href {https://doi.org/10.1103/PhysRevLett.107.073601} {\bibfield  {journal} {\bibinfo  {journal} {Physical Review Letters}\ }\textbf {\bibinfo {volume} {107}},\ \bibinfo {pages} {073601} (\bibinfo {year} {2011})}\BibitemShut {NoStop}%
\bibitem [{\citenamefont {Kannan}\ \emph {et~al.}(2020{\natexlab{a}})\citenamefont {Kannan}, \citenamefont {Campbell}, \citenamefont {Vasconcelos}, \citenamefont {Winik}, \citenamefont {Kim}, \citenamefont {Kjaergaard}, \citenamefont {Krantz}, \citenamefont {Melville}, \citenamefont {Niedzielski}, \citenamefont {Yoder}, \citenamefont {Orlando}, \citenamefont {Gustavsson},\ and\ \citenamefont {Oliver}}]{kannanGeneratingSpatiallyEntangled2020a}%
  \BibitemOpen
  \bibfield  {author} {\bibinfo {author} {\bibfnamefont {B.}~\bibnamefont {Kannan}}, \bibinfo {author} {\bibfnamefont {D.~L.}\ \bibnamefont {Campbell}}, \bibinfo {author} {\bibfnamefont {F.}~\bibnamefont {Vasconcelos}}, \bibinfo {author} {\bibfnamefont {R.}~\bibnamefont {Winik}}, \bibinfo {author} {\bibfnamefont {D.~K.}\ \bibnamefont {Kim}}, \bibinfo {author} {\bibfnamefont {M.}~\bibnamefont {Kjaergaard}}, \bibinfo {author} {\bibfnamefont {P.}~\bibnamefont {Krantz}}, \bibinfo {author} {\bibfnamefont {A.}~\bibnamefont {Melville}}, \bibinfo {author} {\bibfnamefont {B.~M.}\ \bibnamefont {Niedzielski}}, \bibinfo {author} {\bibfnamefont {J.~L.}\ \bibnamefont {Yoder}}, \bibinfo {author} {\bibfnamefont {T.~P.}\ \bibnamefont {Orlando}}, \bibinfo {author} {\bibfnamefont {S.}~\bibnamefont {Gustavsson}},\ and\ \bibinfo {author} {\bibfnamefont {W.~D.}\ \bibnamefont {Oliver}},\ }\bibfield  {title} {\bibinfo {title} {Generating spatially entangled itinerant photons with waveguide quantum electrodynamics},\ }\href
  {https://doi.org/10/gh6jbg} {\bibfield  {journal} {\bibinfo  {journal} {Science Advances}\ }\textbf {\bibinfo {volume} {6}},\ \bibinfo {pages} {eabb8780} (\bibinfo {year} {2020}{\natexlab{a}})}\BibitemShut {NoStop}%
\bibitem [{\citenamefont {Sundaresan}\ \emph {et~al.}(2019)\citenamefont {Sundaresan}, \citenamefont {Lundgren}, \citenamefont {Zhu}, \citenamefont {Gorshkov},\ and\ \citenamefont {Houck}}]{sundaresanInteractingQubitPhotonBound2019}%
  \BibitemOpen
  \bibfield  {author} {\bibinfo {author} {\bibfnamefont {N.~M.}\ \bibnamefont {Sundaresan}}, \bibinfo {author} {\bibfnamefont {R.}~\bibnamefont {Lundgren}}, \bibinfo {author} {\bibfnamefont {G.}~\bibnamefont {Zhu}}, \bibinfo {author} {\bibfnamefont {A.~V.}\ \bibnamefont {Gorshkov}},\ and\ \bibinfo {author} {\bibfnamefont {A.~A.}\ \bibnamefont {Houck}},\ }\bibfield  {title} {\bibinfo {title} {Interacting {{Qubit-Photon Bound States}} with {{Superconducting Circuits}}},\ }\href {https://doi.org/10/gfvs8t} {\bibfield  {journal} {\bibinfo  {journal} {Physical Review X}\ }\textbf {\bibinfo {volume} {9}},\ \bibinfo {pages} {011021} (\bibinfo {year} {2019})}\BibitemShut {NoStop}%
\bibitem [{\citenamefont {Storz}\ \emph {et~al.}(2023)\citenamefont {Storz}, \citenamefont {Sch{\"a}r}, \citenamefont {Kulikov}, \citenamefont {Magnard}, \citenamefont {Kurpiers}, \citenamefont {L{\"u}tolf}, \citenamefont {Walter}, \citenamefont {Copetudo}, \citenamefont {Reuer}, \citenamefont {Akin}, \citenamefont {Besse}, \citenamefont {Gabureac}, \citenamefont {Norris}, \citenamefont {Rosario}, \citenamefont {Martin}, \citenamefont {Martinez}, \citenamefont {Amaya}, \citenamefont {Mitchell}, \citenamefont {Abellan}, \citenamefont {Bancal}, \citenamefont {Sangouard}, \citenamefont {Royer}, \citenamefont {Blais},\ and\ \citenamefont {Wallraff}}]{storzLoopholefreeBellInequality2023}%
  \BibitemOpen
  \bibfield  {author} {\bibinfo {author} {\bibfnamefont {S.}~\bibnamefont {Storz}}, \bibinfo {author} {\bibfnamefont {J.}~\bibnamefont {Sch{\"a}r}}, \bibinfo {author} {\bibfnamefont {A.}~\bibnamefont {Kulikov}}, \bibinfo {author} {\bibfnamefont {P.}~\bibnamefont {Magnard}}, \bibinfo {author} {\bibfnamefont {P.}~\bibnamefont {Kurpiers}}, \bibinfo {author} {\bibfnamefont {J.}~\bibnamefont {L{\"u}tolf}}, \bibinfo {author} {\bibfnamefont {T.}~\bibnamefont {Walter}}, \bibinfo {author} {\bibfnamefont {A.}~\bibnamefont {Copetudo}}, \bibinfo {author} {\bibfnamefont {K.}~\bibnamefont {Reuer}}, \bibinfo {author} {\bibfnamefont {A.}~\bibnamefont {Akin}}, \bibinfo {author} {\bibfnamefont {J.-C.}\ \bibnamefont {Besse}}, \bibinfo {author} {\bibfnamefont {M.}~\bibnamefont {Gabureac}}, \bibinfo {author} {\bibfnamefont {G.~J.}\ \bibnamefont {Norris}}, \bibinfo {author} {\bibfnamefont {A.}~\bibnamefont {Rosario}}, \bibinfo {author} {\bibfnamefont {F.}~\bibnamefont {Martin}}, \bibinfo {author} {\bibfnamefont {J.}~\bibnamefont
  {Martinez}}, \bibinfo {author} {\bibfnamefont {W.}~\bibnamefont {Amaya}}, \bibinfo {author} {\bibfnamefont {M.~W.}\ \bibnamefont {Mitchell}}, \bibinfo {author} {\bibfnamefont {C.}~\bibnamefont {Abellan}}, \bibinfo {author} {\bibfnamefont {J.-D.}\ \bibnamefont {Bancal}}, \bibinfo {author} {\bibfnamefont {N.}~\bibnamefont {Sangouard}}, \bibinfo {author} {\bibfnamefont {B.}~\bibnamefont {Royer}}, \bibinfo {author} {\bibfnamefont {A.}~\bibnamefont {Blais}},\ and\ \bibinfo {author} {\bibfnamefont {A.}~\bibnamefont {Wallraff}},\ }\bibfield  {title} {\bibinfo {title} {Loophole-free {{Bell}} inequality violation with superconducting circuits},\ }\href {https://doi.org/10.1038/s41586-023-05885-0} {\bibfield  {journal} {\bibinfo  {journal} {Nature}\ }\textbf {\bibinfo {volume} {617}},\ \bibinfo {pages} {265} (\bibinfo {year} {2023})}\BibitemShut {NoStop}%
\bibitem [{\citenamefont {Kannan}\ \emph {et~al.}(2020{\natexlab{b}})\citenamefont {Kannan}, \citenamefont {Ruckriegel}, \citenamefont {Campbell}, \citenamefont {Frisk~Kockum}, \citenamefont {Braum{\"u}ller}, \citenamefont {Kim}, \citenamefont {Kjaergaard}, \citenamefont {Krantz}, \citenamefont {Melville}, \citenamefont {Niedzielski}, \citenamefont {Veps{\"a}l{\"a}inen}, \citenamefont {Winik}, \citenamefont {Yoder}, \citenamefont {Nori}, \citenamefont {Orlando}, \citenamefont {Gustavsson},\ and\ \citenamefont {Oliver}}]{kannanWaveguideQuantumElectrodynamics2020}%
  \BibitemOpen
  \bibfield  {author} {\bibinfo {author} {\bibfnamefont {B.}~\bibnamefont {Kannan}}, \bibinfo {author} {\bibfnamefont {M.~J.}\ \bibnamefont {Ruckriegel}}, \bibinfo {author} {\bibfnamefont {D.~L.}\ \bibnamefont {Campbell}}, \bibinfo {author} {\bibfnamefont {A.}~\bibnamefont {Frisk~Kockum}}, \bibinfo {author} {\bibfnamefont {J.}~\bibnamefont {Braum{\"u}ller}}, \bibinfo {author} {\bibfnamefont {D.~K.}\ \bibnamefont {Kim}}, \bibinfo {author} {\bibfnamefont {M.}~\bibnamefont {Kjaergaard}}, \bibinfo {author} {\bibfnamefont {P.}~\bibnamefont {Krantz}}, \bibinfo {author} {\bibfnamefont {A.}~\bibnamefont {Melville}}, \bibinfo {author} {\bibfnamefont {B.~M.}\ \bibnamefont {Niedzielski}}, \bibinfo {author} {\bibfnamefont {A.}~\bibnamefont {Veps{\"a}l{\"a}inen}}, \bibinfo {author} {\bibfnamefont {R.}~\bibnamefont {Winik}}, \bibinfo {author} {\bibfnamefont {J.~L.}\ \bibnamefont {Yoder}}, \bibinfo {author} {\bibfnamefont {F.}~\bibnamefont {Nori}}, \bibinfo {author} {\bibfnamefont {T.~P.}\ \bibnamefont {Orlando}}, \bibinfo
  {author} {\bibfnamefont {S.}~\bibnamefont {Gustavsson}},\ and\ \bibinfo {author} {\bibfnamefont {W.~D.}\ \bibnamefont {Oliver}},\ }\bibfield  {title} {\bibinfo {title} {Waveguide quantum electrodynamics with superconducting artificial giant atoms},\ }\href {https://doi.org/10.1038/s41586-020-2529-9} {\bibfield  {journal} {\bibinfo  {journal} {Nature}\ }\textbf {\bibinfo {volume} {583}},\ \bibinfo {pages} {775} (\bibinfo {year} {2020}{\natexlab{b}})}\BibitemShut {NoStop}%
\bibitem [{\citenamefont {Rosario~Hamann}\ \emph {et~al.}(2018)\citenamefont {Rosario~Hamann}, \citenamefont {M{\"u}ller}, \citenamefont {Jerger}, \citenamefont {Zanner}, \citenamefont {Combes}, \citenamefont {Pletyukhov}, \citenamefont {Weides}, \citenamefont {Stace},\ and\ \citenamefont {Fedorov}}]{rosariohamannNonreciprocityRealizedQuantum2018}%
  \BibitemOpen
  \bibfield  {author} {\bibinfo {author} {\bibfnamefont {A.}~\bibnamefont {Rosario~Hamann}}, \bibinfo {author} {\bibfnamefont {C.}~\bibnamefont {M{\"u}ller}}, \bibinfo {author} {\bibfnamefont {M.}~\bibnamefont {Jerger}}, \bibinfo {author} {\bibfnamefont {M.}~\bibnamefont {Zanner}}, \bibinfo {author} {\bibfnamefont {J.}~\bibnamefont {Combes}}, \bibinfo {author} {\bibfnamefont {M.}~\bibnamefont {Pletyukhov}}, \bibinfo {author} {\bibfnamefont {M.}~\bibnamefont {Weides}}, \bibinfo {author} {\bibfnamefont {T.~M.}\ \bibnamefont {Stace}},\ and\ \bibinfo {author} {\bibfnamefont {A.}~\bibnamefont {Fedorov}},\ }\bibfield  {title} {\bibinfo {title} {Nonreciprocity {{Realized}} with {{Quantum Nonlinearity}}},\ }\href {https://doi.org/10/gd6gr6} {\bibfield  {journal} {\bibinfo  {journal} {Physical Review Letters}\ }\textbf {\bibinfo {volume} {121}},\ \bibinfo {pages} {123601} (\bibinfo {year} {2018})}\BibitemShut {NoStop}%
\bibitem [{\citenamefont {Kannan}\ \emph {et~al.}(2023)\citenamefont {Kannan}, \citenamefont {Almanakly}, \citenamefont {Sung}, \citenamefont {Di~Paolo}, \citenamefont {Rower}, \citenamefont {Braum{\"u}ller}, \citenamefont {Melville}, \citenamefont {Niedzielski}, \citenamefont {Karamlou}, \citenamefont {Serniak}, \citenamefont {Veps{\"a}l{\"a}inen}, \citenamefont {Schwartz}, \citenamefont {Yoder}, \citenamefont {Winik}, \citenamefont {Wang}, \citenamefont {Orlando}, \citenamefont {Gustavsson}, \citenamefont {Grover},\ and\ \citenamefont {Oliver}}]{kannanOndemandDirectionalMicrowave2023}%
  \BibitemOpen
  \bibfield  {author} {\bibinfo {author} {\bibfnamefont {B.}~\bibnamefont {Kannan}}, \bibinfo {author} {\bibfnamefont {A.}~\bibnamefont {Almanakly}}, \bibinfo {author} {\bibfnamefont {Y.}~\bibnamefont {Sung}}, \bibinfo {author} {\bibfnamefont {A.}~\bibnamefont {Di~Paolo}}, \bibinfo {author} {\bibfnamefont {D.~A.}\ \bibnamefont {Rower}}, \bibinfo {author} {\bibfnamefont {J.}~\bibnamefont {Braum{\"u}ller}}, \bibinfo {author} {\bibfnamefont {A.}~\bibnamefont {Melville}}, \bibinfo {author} {\bibfnamefont {B.~M.}\ \bibnamefont {Niedzielski}}, \bibinfo {author} {\bibfnamefont {A.}~\bibnamefont {Karamlou}}, \bibinfo {author} {\bibfnamefont {K.}~\bibnamefont {Serniak}}, \bibinfo {author} {\bibfnamefont {A.}~\bibnamefont {Veps{\"a}l{\"a}inen}}, \bibinfo {author} {\bibfnamefont {M.~E.}\ \bibnamefont {Schwartz}}, \bibinfo {author} {\bibfnamefont {J.~L.}\ \bibnamefont {Yoder}}, \bibinfo {author} {\bibfnamefont {R.}~\bibnamefont {Winik}}, \bibinfo {author} {\bibfnamefont {J.~I.-J.}\ \bibnamefont {Wang}}, \bibinfo {author}
  {\bibfnamefont {T.~P.}\ \bibnamefont {Orlando}}, \bibinfo {author} {\bibfnamefont {S.}~\bibnamefont {Gustavsson}}, \bibinfo {author} {\bibfnamefont {J.~A.}\ \bibnamefont {Grover}},\ and\ \bibinfo {author} {\bibfnamefont {W.~D.}\ \bibnamefont {Oliver}},\ }\bibfield  {title} {\bibinfo {title} {On-demand directional microwave photon emission using waveguide quantum electrodynamics},\ }\href {https://doi.org/10.1038/s41567-022-01869-5} {\bibfield  {journal} {\bibinfo  {journal} {Nature Physics}\ }\textbf {\bibinfo {volume} {19}},\ \bibinfo {pages} {394} (\bibinfo {year} {2023})}\BibitemShut {NoStop}%
\bibitem [{\citenamefont {Joshi}\ \emph {et~al.}(2023)\citenamefont {Joshi}, \citenamefont {Yang},\ and\ \citenamefont {Mirhosseini}}]{joshiResonanceFluorescenceChiral2023}%
  \BibitemOpen
  \bibfield  {author} {\bibinfo {author} {\bibfnamefont {C.}~\bibnamefont {Joshi}}, \bibinfo {author} {\bibfnamefont {F.}~\bibnamefont {Yang}},\ and\ \bibinfo {author} {\bibfnamefont {M.}~\bibnamefont {Mirhosseini}},\ }\bibfield  {title} {\bibinfo {title} {Resonance {{Fluorescence}} of a {{Chiral Artificial Atom}}},\ }\href {https://doi.org/10.1103/PhysRevX.13.021039} {\bibfield  {journal} {\bibinfo  {journal} {Physical Review X}\ }\textbf {\bibinfo {volume} {13}},\ \bibinfo {pages} {021039} (\bibinfo {year} {2023})}\BibitemShut {NoStop}%
\bibitem [{\citenamefont {Kim}\ \emph {et~al.}(2021)\citenamefont {Kim}, \citenamefont {Zhang}, \citenamefont {Ferreira}, \citenamefont {Banker}, \citenamefont {Iverson}, \citenamefont {Sipahigil}, \citenamefont {Bello}, \citenamefont {{Gonz{\'a}lez-Tudela}}, \citenamefont {Mirhosseini},\ and\ \citenamefont {Painter}}]{kimQuantumElectrodynamicsTopological2021a}%
  \BibitemOpen
  \bibfield  {author} {\bibinfo {author} {\bibfnamefont {E.}~\bibnamefont {Kim}}, \bibinfo {author} {\bibfnamefont {X.}~\bibnamefont {Zhang}}, \bibinfo {author} {\bibfnamefont {V.~S.}\ \bibnamefont {Ferreira}}, \bibinfo {author} {\bibfnamefont {J.}~\bibnamefont {Banker}}, \bibinfo {author} {\bibfnamefont {J.~K.}\ \bibnamefont {Iverson}}, \bibinfo {author} {\bibfnamefont {A.}~\bibnamefont {Sipahigil}}, \bibinfo {author} {\bibfnamefont {M.}~\bibnamefont {Bello}}, \bibinfo {author} {\bibfnamefont {A.}~\bibnamefont {{Gonz{\'a}lez-Tudela}}}, \bibinfo {author} {\bibfnamefont {M.}~\bibnamefont {Mirhosseini}},\ and\ \bibinfo {author} {\bibfnamefont {O.}~\bibnamefont {Painter}},\ }\bibfield  {title} {\bibinfo {title} {Quantum {{Electrodynamics}} in a {{Topological Waveguide}}},\ }\href {https://doi.org/10/gh2n45} {\bibfield  {journal} {\bibinfo  {journal} {Physical Review X}\ }\textbf {\bibinfo {volume} {11}},\ \bibinfo {pages} {011015} (\bibinfo {year} {2021})}\BibitemShut {NoStop}%
\bibitem [{\citenamefont {{Gonz{\'a}lez-Tudela}}\ \emph {et~al.}(2015)\citenamefont {{Gonz{\'a}lez-Tudela}}, \citenamefont {Hung}, \citenamefont {Chang}, \citenamefont {Cirac},\ and\ \citenamefont {Kimble}}]{gonzalez-tudelaSubwavelengthVacuumLattices2015}%
  \BibitemOpen
  \bibfield  {author} {\bibinfo {author} {\bibfnamefont {A.}~\bibnamefont {{Gonz{\'a}lez-Tudela}}}, \bibinfo {author} {\bibfnamefont {C.-L.}\ \bibnamefont {Hung}}, \bibinfo {author} {\bibfnamefont {D.~E.}\ \bibnamefont {Chang}}, \bibinfo {author} {\bibfnamefont {J.~I.}\ \bibnamefont {Cirac}},\ and\ \bibinfo {author} {\bibfnamefont {H.~J.}\ \bibnamefont {Kimble}},\ }\bibfield  {title} {\bibinfo {title} {Subwavelength vacuum lattices and atom--atom interactions in two-dimensional photonic crystals},\ }\href {https://doi.org/10.1038/nphoton.2015.54} {\bibfield  {journal} {\bibinfo  {journal} {Nature Photonics}\ }\textbf {\bibinfo {volume} {9}},\ \bibinfo {pages} {320} (\bibinfo {year} {2015})}\BibitemShut {NoStop}%
\bibitem [{\citenamefont {Lodahl}\ \emph {et~al.}(2004)\citenamefont {Lodahl}, \citenamefont {{Floris van Driel}}, \citenamefont {Nikolaev}, \citenamefont {Irman}, \citenamefont {Overgaag}, \citenamefont {Vanmaekelbergh},\ and\ \citenamefont {Vos}}]{lodahlControllingDynamicsSpontaneous2004}%
  \BibitemOpen
  \bibfield  {author} {\bibinfo {author} {\bibfnamefont {P.}~\bibnamefont {Lodahl}}, \bibinfo {author} {\bibfnamefont {A.}~\bibnamefont {{Floris van Driel}}}, \bibinfo {author} {\bibfnamefont {I.~S.}\ \bibnamefont {Nikolaev}}, \bibinfo {author} {\bibfnamefont {A.}~\bibnamefont {Irman}}, \bibinfo {author} {\bibfnamefont {K.}~\bibnamefont {Overgaag}}, \bibinfo {author} {\bibfnamefont {D.}~\bibnamefont {Vanmaekelbergh}},\ and\ \bibinfo {author} {\bibfnamefont {W.~L.}\ \bibnamefont {Vos}},\ }\bibfield  {title} {\bibinfo {title} {Controlling the dynamics of spontaneous emission from quantum dots by photonic crystals},\ }\href {https://doi.org/10.1038/nature02772} {\bibfield  {journal} {\bibinfo  {journal} {Nature}\ }\textbf {\bibinfo {volume} {430}},\ \bibinfo {pages} {654} (\bibinfo {year} {2004})}\BibitemShut {NoStop}%
\bibitem [{\citenamefont {Boyd}\ \emph {et~al.}(2009)\citenamefont {Boyd}, \citenamefont {Lukishova}, \citenamefont {Shen},\ and\ \citenamefont {Ascheron}}]{boydSelffocusingPresentFundamentals2009}%
  \BibitemOpen
  \bibinfo {editor} {\bibfnamefont {R.~W.}\ \bibnamefont {Boyd}}, \bibinfo {editor} {\bibfnamefont {S.~G.}\ \bibnamefont {Lukishova}}, \bibinfo {editor} {\bibfnamefont {Y.}~\bibnamefont {Shen}},\ and\ \bibinfo {editor} {\bibfnamefont {C.}~\bibnamefont {Ascheron}},\ eds.,\ \href {https://doi.org/10.1007/978-0-387-34727-1} {\emph {\bibinfo {title} {Self-Focusing: {{Past}} and {{Present}}: {{Fundamentals}} and {{Prospects}}}}},\ \bibinfo {series} {Topics in {{Applied Physics}}}, Vol.\ \bibinfo {volume} {114}\ (\bibinfo  {publisher} {Springer},\ \bibinfo {address} {New York, NY},\ \bibinfo {year} {2009})\BibitemShut {NoStop}%
\bibitem [{\citenamefont {Krausz}\ and\ \citenamefont {Ivanov}(2009)}]{krauszAttosecondPhysics2009}%
  \BibitemOpen
  \bibfield  {author} {\bibinfo {author} {\bibfnamefont {F.}~\bibnamefont {Krausz}}\ and\ \bibinfo {author} {\bibfnamefont {M.}~\bibnamefont {Ivanov}},\ }\bibfield  {title} {\bibinfo {title} {Attosecond physics},\ }\href {https://doi.org/10.1103/RevModPhys.81.163} {\bibfield  {journal} {\bibinfo  {journal} {Reviews of Modern Physics}\ }\textbf {\bibinfo {volume} {81}},\ \bibinfo {pages} {163} (\bibinfo {year} {2009})}\BibitemShut {NoStop}%
\bibitem [{\citenamefont {Casulleras}\ \emph {et~al.}(2021)\citenamefont {Casulleras}, \citenamefont {{Gonzalez-Ballestero}}, \citenamefont {Maurer}, \citenamefont {{Garc{\'i}a-Ripoll}},\ and\ \citenamefont {{Romero-Isart}}}]{casullerasRemoteIndividualAddressing2021}%
  \BibitemOpen
  \bibfield  {author} {\bibinfo {author} {\bibfnamefont {S.}~\bibnamefont {Casulleras}}, \bibinfo {author} {\bibfnamefont {C.}~\bibnamefont {{Gonzalez-Ballestero}}}, \bibinfo {author} {\bibfnamefont {P.}~\bibnamefont {Maurer}}, \bibinfo {author} {\bibfnamefont {J.~J.}\ \bibnamefont {{Garc{\'i}a-Ripoll}}},\ and\ \bibinfo {author} {\bibfnamefont {O.}~\bibnamefont {{Romero-Isart}}},\ }\bibfield  {title} {\bibinfo {title} {Remote {{Individual Addressing}} of {{Quantum Emitters}} with {{Chirped Pulses}}},\ }\href {https://doi.org/10.1103/PhysRevLett.126.103602} {\bibfield  {journal} {\bibinfo  {journal} {Physical Review Letters}\ }\textbf {\bibinfo {volume} {126}},\ \bibinfo {pages} {103602} (\bibinfo {year} {2021})}\BibitemShut {NoStop}%
\bibitem [{\citenamefont {Koch}\ \emph {et~al.}(2007)\citenamefont {Koch}, \citenamefont {Yu}, \citenamefont {Gambetta}, \citenamefont {Houck}, \citenamefont {Schuster}, \citenamefont {Majer}, \citenamefont {Blais}, \citenamefont {Devoret}, \citenamefont {Girvin},\ and\ \citenamefont {Schoelkopf}}]{kochChargeinsensitiveQubitDesign2007b}%
  \BibitemOpen
  \bibfield  {author} {\bibinfo {author} {\bibfnamefont {J.}~\bibnamefont {Koch}}, \bibinfo {author} {\bibfnamefont {T.~M.}\ \bibnamefont {Yu}}, \bibinfo {author} {\bibfnamefont {J.}~\bibnamefont {Gambetta}}, \bibinfo {author} {\bibfnamefont {A.~A.}\ \bibnamefont {Houck}}, \bibinfo {author} {\bibfnamefont {D.~I.}\ \bibnamefont {Schuster}}, \bibinfo {author} {\bibfnamefont {J.}~\bibnamefont {Majer}}, \bibinfo {author} {\bibfnamefont {A.}~\bibnamefont {Blais}}, \bibinfo {author} {\bibfnamefont {M.~H.}\ \bibnamefont {Devoret}}, \bibinfo {author} {\bibfnamefont {S.~M.}\ \bibnamefont {Girvin}},\ and\ \bibinfo {author} {\bibfnamefont {R.~J.}\ \bibnamefont {Schoelkopf}},\ }\bibfield  {title} {\bibinfo {title} {Charge-insensitive qubit design derived from the {{Cooper}} pair box},\ }\href {https://doi.org/10/b5bt9m} {\bibfield  {journal} {\bibinfo  {journal} {Physical Review A}\ }\textbf {\bibinfo {volume} {76}},\ \bibinfo {pages} {042319} (\bibinfo {year} {2007})}\BibitemShut {NoStop}%
\bibitem [{\citenamefont {Zoepfl}\ \emph {et~al.}(2017)\citenamefont {Zoepfl}, \citenamefont {Muppalla}, \citenamefont {Schneider}, \citenamefont {Kasemann}, \citenamefont {Partel},\ and\ \citenamefont {Kirchmair}}]{zoepflCharacterizationLowLoss2017}%
  \BibitemOpen
  \bibfield  {author} {\bibinfo {author} {\bibfnamefont {D.}~\bibnamefont {Zoepfl}}, \bibinfo {author} {\bibfnamefont {P.~R.}\ \bibnamefont {Muppalla}}, \bibinfo {author} {\bibfnamefont {C.~M.~F.}\ \bibnamefont {Schneider}}, \bibinfo {author} {\bibfnamefont {S.}~\bibnamefont {Kasemann}}, \bibinfo {author} {\bibfnamefont {S.}~\bibnamefont {Partel}},\ and\ \bibinfo {author} {\bibfnamefont {G.}~\bibnamefont {Kirchmair}},\ }\bibfield  {title} {\bibinfo {title} {Characterization of low loss microstrip resonators as a building block for circuit {{QED}} in a {{3D}} waveguide},\ }\href {https://doi.org/10.1063/1.4992070} {\bibfield  {journal} {\bibinfo  {journal} {AIP Advances}\ }\textbf {\bibinfo {volume} {7}},\ \bibinfo {pages} {085118} (\bibinfo {year} {2017})}\BibitemShut {NoStop}%
\bibitem [{\citenamefont {Lalumi{\`e}re}\ \emph {et~al.}(2013)\citenamefont {Lalumi{\`e}re}, \citenamefont {Sanders}, \citenamefont {{van Loo}}, \citenamefont {Fedorov}, \citenamefont {Wallraff},\ and\ \citenamefont {Blais}}]{lalumiereInputoutputTheoryWaveguide2013b}%
  \BibitemOpen
  \bibfield  {author} {\bibinfo {author} {\bibfnamefont {K.}~\bibnamefont {Lalumi{\`e}re}}, \bibinfo {author} {\bibfnamefont {B.~C.}\ \bibnamefont {Sanders}}, \bibinfo {author} {\bibfnamefont {A.~F.}\ \bibnamefont {{van Loo}}}, \bibinfo {author} {\bibfnamefont {A.}~\bibnamefont {Fedorov}}, \bibinfo {author} {\bibfnamefont {A.}~\bibnamefont {Wallraff}},\ and\ \bibinfo {author} {\bibfnamefont {A.}~\bibnamefont {Blais}},\ }\bibfield  {title} {\bibinfo {title} {Input-output theory for waveguide {{QED}} with an ensemble of inhomogeneous atoms},\ }\href {https://doi.org/10/gf2s48} {\bibfield  {journal} {\bibinfo  {journal} {Physical Review A}\ }\textbf {\bibinfo {volume} {88}},\ \bibinfo {pages} {043806} (\bibinfo {year} {2013})}\BibitemShut {NoStop}%
\bibitem [{\citenamefont {Brehm}\ \emph {et~al.}(2021)\citenamefont {Brehm}, \citenamefont {Poddubny}, \citenamefont {Stehli}, \citenamefont {Wolz}, \citenamefont {Rotzinger},\ and\ \citenamefont {Ustinov}}]{brehmWaveguideBandgapEngineering2021}%
  \BibitemOpen
  \bibfield  {author} {\bibinfo {author} {\bibfnamefont {J.~D.}\ \bibnamefont {Brehm}}, \bibinfo {author} {\bibfnamefont {A.~N.}\ \bibnamefont {Poddubny}}, \bibinfo {author} {\bibfnamefont {A.}~\bibnamefont {Stehli}}, \bibinfo {author} {\bibfnamefont {T.}~\bibnamefont {Wolz}}, \bibinfo {author} {\bibfnamefont {H.}~\bibnamefont {Rotzinger}},\ and\ \bibinfo {author} {\bibfnamefont {A.~V.}\ \bibnamefont {Ustinov}},\ }\bibfield  {title} {\bibinfo {title} {Waveguide bandgap engineering with an array of superconducting qubits},\ }\href {https://doi.org/10/ghzxcz} {\bibfield  {journal} {\bibinfo  {journal} {npj Quantum Materials}\ }\textbf {\bibinfo {volume} {6}},\ \bibinfo {pages} {1} (\bibinfo {year} {2021})}\BibitemShut {NoStop}%
\bibitem [{\citenamefont {Poshakinskiy}\ \emph {et~al.}(2021)\citenamefont {Poshakinskiy}, \citenamefont {Zhong},\ and\ \citenamefont {Poddubny}}]{poshakinskiyQuantumChaosDriven2021}%
  \BibitemOpen
  \bibfield  {author} {\bibinfo {author} {\bibfnamefont {A.~V.}\ \bibnamefont {Poshakinskiy}}, \bibinfo {author} {\bibfnamefont {J.}~\bibnamefont {Zhong}},\ and\ \bibinfo {author} {\bibfnamefont {A.~N.}\ \bibnamefont {Poddubny}},\ }\bibfield  {title} {\bibinfo {title} {Quantum {{Chaos Driven}} by {{Long-Range Waveguide-Mediated Interactions}}},\ }\href {https://doi.org/10.1103/PhysRevLett.126.203602} {\bibfield  {journal} {\bibinfo  {journal} {Physical Review Letters}\ }\textbf {\bibinfo {volume} {126}},\ \bibinfo {pages} {203602} (\bibinfo {year} {2021})}\BibitemShut {NoStop}%
\bibitem [{\citenamefont {Manucharyan}\ \emph {et~al.}(2009)\citenamefont {Manucharyan}, \citenamefont {Koch}, \citenamefont {Glazman},\ and\ \citenamefont {Devoret}}]{manucharyanFluxoniumSingleCooperPair2009b}%
  \BibitemOpen
  \bibfield  {author} {\bibinfo {author} {\bibfnamefont {V.~E.}\ \bibnamefont {Manucharyan}}, \bibinfo {author} {\bibfnamefont {J.}~\bibnamefont {Koch}}, \bibinfo {author} {\bibfnamefont {L.~I.}\ \bibnamefont {Glazman}},\ and\ \bibinfo {author} {\bibfnamefont {M.~H.}\ \bibnamefont {Devoret}},\ }\bibfield  {title} {\bibinfo {title} {Fluxonium: {{Single Cooper-Pair Circuit Free}} of {{Charge Offsets}}},\ }\href {https://doi.org/10.1126/science.1175552} {\bibfield  {journal} {\bibinfo  {journal} {Science}\ }\textbf {\bibinfo {volume} {326}},\ \bibinfo {pages} {113} (\bibinfo {year} {2009})}\BibitemShut {NoStop}%
\bibitem [{\citenamefont {Wolfowicz}\ \emph {et~al.}(2021)\citenamefont {Wolfowicz}, \citenamefont {Heremans}, \citenamefont {Anderson}, \citenamefont {Kanai}, \citenamefont {Seo}, \citenamefont {Gali}, \citenamefont {Galli},\ and\ \citenamefont {Awschalom}}]{wolfowiczQuantumGuidelinesSolidstate2021}%
  \BibitemOpen
  \bibfield  {author} {\bibinfo {author} {\bibfnamefont {G.}~\bibnamefont {Wolfowicz}}, \bibinfo {author} {\bibfnamefont {F.~J.}\ \bibnamefont {Heremans}}, \bibinfo {author} {\bibfnamefont {C.~P.}\ \bibnamefont {Anderson}}, \bibinfo {author} {\bibfnamefont {S.}~\bibnamefont {Kanai}}, \bibinfo {author} {\bibfnamefont {H.}~\bibnamefont {Seo}}, \bibinfo {author} {\bibfnamefont {A.}~\bibnamefont {Gali}}, \bibinfo {author} {\bibfnamefont {G.}~\bibnamefont {Galli}},\ and\ \bibinfo {author} {\bibfnamefont {D.~D.}\ \bibnamefont {Awschalom}},\ }\bibfield  {title} {\bibinfo {title} {Quantum guidelines for solid-state spin defects},\ }\href {https://doi.org/10.1038/s41578-021-00306-y} {\bibfield  {journal} {\bibinfo  {journal} {Nature Reviews Materials}\ }\textbf {\bibinfo {volume} {6}},\ \bibinfo {pages} {906} (\bibinfo {year} {2021})}\BibitemShut {NoStop}%
\bibitem [{\citenamefont {Boissonneault}\ \emph {et~al.}(2008)\citenamefont {Boissonneault}, \citenamefont {Gambetta},\ and\ \citenamefont {Blais}}]{boissonneaultNonlinearDispersiveRegime2008}%
  \BibitemOpen
  \bibfield  {author} {\bibinfo {author} {\bibfnamefont {M.}~\bibnamefont {Boissonneault}}, \bibinfo {author} {\bibfnamefont {J.~M.}\ \bibnamefont {Gambetta}},\ and\ \bibinfo {author} {\bibfnamefont {A.}~\bibnamefont {Blais}},\ }\bibfield  {title} {\bibinfo {title} {Nonlinear dispersive regime of cavity {{QED}}: {{The}} dressed dephasing model},\ }\href {https://doi.org/10.1103/PhysRevA.77.060305} {\bibfield  {journal} {\bibinfo  {journal} {Physical Review A}\ }\textbf {\bibinfo {volume} {77}},\ \bibinfo {pages} {060305} (\bibinfo {year} {2008})}\BibitemShut {NoStop}%
\bibitem [{\citenamefont {Dalmonte}\ \emph {et~al.}(2015)\citenamefont {Dalmonte}, \citenamefont {Mirzaei}, \citenamefont {Muppalla}, \citenamefont {Marcos}, \citenamefont {Zoller},\ and\ \citenamefont {Kirchmair}}]{dalmonteRealizingDipolarSpin2015}%
  \BibitemOpen
  \bibfield  {author} {\bibinfo {author} {\bibfnamefont {M.}~\bibnamefont {Dalmonte}}, \bibinfo {author} {\bibfnamefont {S.~I.}\ \bibnamefont {Mirzaei}}, \bibinfo {author} {\bibfnamefont {P.~R.}\ \bibnamefont {Muppalla}}, \bibinfo {author} {\bibfnamefont {D.}~\bibnamefont {Marcos}}, \bibinfo {author} {\bibfnamefont {P.}~\bibnamefont {Zoller}},\ and\ \bibinfo {author} {\bibfnamefont {G.}~\bibnamefont {Kirchmair}},\ }\bibfield  {title} {\bibinfo {title} {Realizing dipolar spin models with arrays of superconducting qubits},\ }\href {https://doi.org/10/ggz3qt} {\bibfield  {journal} {\bibinfo  {journal} {Physical Review B}\ }\textbf {\bibinfo {volume} {92}},\ \bibinfo {pages} {174507} (\bibinfo {year} {2015})}\BibitemShut {NoStop}%
\bibitem [{\citenamefont {Liu}\ and\ \citenamefont {Houck}(2017)}]{liuQuantumElectrodynamicsPhotonic2017b}%
  \BibitemOpen
  \bibfield  {author} {\bibinfo {author} {\bibfnamefont {Y.}~\bibnamefont {Liu}}\ and\ \bibinfo {author} {\bibfnamefont {A.~A.}\ \bibnamefont {Houck}},\ }\bibfield  {title} {\bibinfo {title} {Quantum electrodynamics near a photonic bandgap},\ }\href {https://doi.org/10.1038/nphys3834} {\bibfield  {journal} {\bibinfo  {journal} {Nature Physics}\ }\textbf {\bibinfo {volume} {13}},\ \bibinfo {pages} {48} (\bibinfo {year} {2017})}\BibitemShut {NoStop}%
\end{thebibliography}%
\onecolumngrid
\appendix
\beginsupplement

\newpage
\section{SUPPLEMENTARY MATERIAL}

\subsection{Theory}
In this section, we describe the impact of dispersion on the propagation of a chirped pulse traveling inside a rectangular waveguide and its interaction with a single quantum emitter, following Ref.~\cite{casullerasRemoteIndividualAddressing2021}. Let us write the classical electrical field associated to a Gaussian pulse in k-vector space, given by 
\begin{equation}
\tilde{E}(k) \propto \exp \left( -\frac{\sigma_f^2(k-k_0)^2}{2} \right),
\label{eq:pulse_def_annex}
\end{equation}
where $\sigma_f$ is the spot size and $k_0$ is the central k-vector of the pulse. To perform the experiment, pulse frequencies near the cutoff of the rectangular waveguide, which are not well coupled to the waveguide (see~\autoref{fig:Fig1}
a), are removed.  Moreover, these frequencies travel very slowly and thus create long AWG waveforms, causing numerical instability. To avoid this, the pulse definition in~\autoref{eq:pulse_def_annex} is modified with a smooth high pass filter, that is,
\begin{equation}
\tilde{E}_2(k) \propto \exp \left( -\frac{\sigma_f^2(k-k_0)^2}{2} \right)\exp \left( \frac{-0.01\times k_c^2}{k^2} \right),
\end{equation}
where the high pass has a 3~dB point of 6.605~GHz (compared to 6.557~GHz for the waveguide cutoff) and does not significantly change the shape of the pulse.

The temporal behavior of the pulse at a given position $z=d_f$ can then be computed using the dispersion relation $\omega(k)$ of the waveguide, as
\begin{equation}
    E_f(t) =E_0\textrm{Re} \left[ \int_{0}^{\infty} \textrm{d}k \tilde{E}_2(k) \exp \left(i \omega(k) t\right) \right],
    \label{eq:pulse_definition_appendix}
\end{equation}
where $\omega(k) = c\sqrt{k^2+k_c^2}$ and $k_c$ is the cutoff k-vector of the rectangular waveguide and  $E_0$ is the amplitude of the pulse.  Here, the cutoff k-vector is linked to the cutoff frequency of the rectangular waveguide through the relation $k_c = \omega_c/c$, where $c$ is the speed of light in the medium (in our case vacuum). The  dispersion relation  is used to compute the central k-vector of pulse, given by  $k_0=\sqrt{\omega_{ge}^2/c^2-k_c^2}$, where $\omega_{ge}$ is the frequency of the transition between the ground state and first excited state of the transmons. Note that the extension of this theoretical description to other systems can be done by changing the dispersion relation $\omega(k)$.

The time-dependent electric field in~\autoref{eq:pulse_definition_appendix} corresponds to the pulse at the focal point, where all the k-components of the pulse are in phase and the maximum amplitude is present. The equation to back-propagate the pulse to arbitrary locations along the waveguide reads
\begin{equation}
    E(t, z, d_f) = \frac{1}{2\pi} \int_{-\infty}^\infty \textrm{d}\omega e^{\textrm{i} \omega t} \left[
    \int_{-\infty}^\infty \textrm{d}t^\prime e^{-\textrm{i} \omega t^\prime} E_f(t^\prime) \right]
    e^{\textrm{i } \textrm{sgn}(\omega) k(\omega) (z-d_f)},
    \label{eq:ap_prop_2}
\end{equation}
where $k(\omega) = \sqrt{\omega^2-\omega_c^2}/c$  and the sign function ensures that $E(t,z,d_f)$ is real. Physically, the sign function corresponds to a wave which propagates only in one direction (towards increasing $z$).  The input of the rectangular waveguide is used as the origin of the $z$ axis. Therefore, the input waveform is computed with~\autoref{eq:ap_prop_2} setting $z=0$.

To gain an intuition on the interaction between a quantum emitter and the self-focusing pulse described by~\autoref{eq:ap_prop_2}, we perform two approximations. First, we consider the quantum emitter to be a two-level system, instead of a transmon with multiple levels of energy. Secondly, we perform the rotating wave approximation, which implies that the Rabi frequency (proportional to the field amplitude) is small compared to the qubit frequency. In that case, the Hamiltonian that describes the qubit dynamics can be written in the rotating frame as
\begin{equation}
    \frac{\hat{H}_{LZ}}{\hbar} = \Delta(t,z) \hat{\sigma}_z + \frac{g(t,z)}{2}\left(\hat{\sigma}_+ + \hat{\sigma}_-\right),
\end{equation}
where $\Delta(t,z)$ is the qubit detuning, $g(z,t)$ is the Rabi coupling and $z$ is the position of the qubit. Both functions $\Delta(t,z)$ and $g(z,t)$ can be computed from the analytical representation of the pulse electrical field $E(t,z) = A(t,z) \exp(i\phi(t,z))$, where $A(t,z)$ is the instantaneous amplitude and $\phi(t,z)$ is the instantaneous phase of the pulse. In particular, the qubit detuning is given by $\Delta(t,z) = \omega_q - \frac{\partial \phi}{\partial t}(t,z)$ where $\omega_q$ is the qubit frequency and $\frac{\partial \phi}{\partial t}(t,z)$ is the instantaneous frequency of the pulse. The Rabi coupling is given by $g(t,z=z) = d_{eg} A(t,z)$ where $d_{eg}$ is the electric dipole moment of the qubit, corresponding to the coupling of the first transition of the qubit to the waveguide.

\begin{figure*}[ht]
    \centering
    \includegraphics[width=1\linewidth]{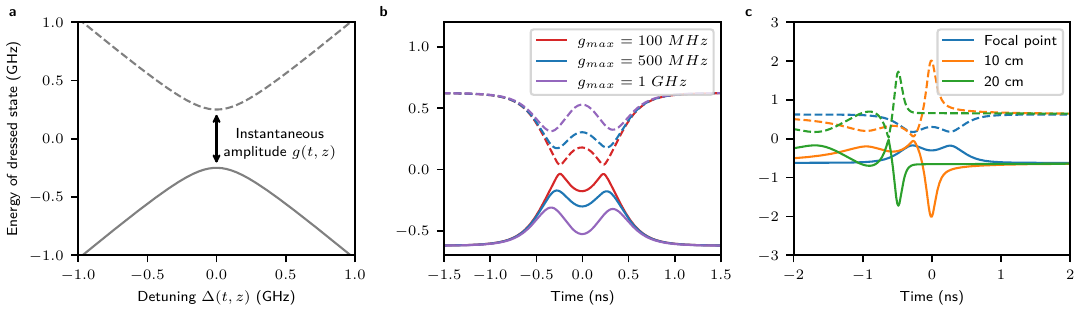}
    \caption{\textbf{Landau-Zener explanation of the qubit dynamics.} \textbf{a} Energy of the dressed states of the qubit in the Landau-Zener picture as a function of the detuning and instantaneous Rabi coupling amplitude. \textbf{b} Energy of the dressed states as a function of time at the focal point for different pulse amplitudes ($g_{max} = \textrm{max}(g(t,z_f))$. \textbf{c} Energy of the dressed states at the focal point, and at 10~cm and 20~cm before the focal point.}
    \label{fig:sup_detuning_vs_amplitude_LZ}
\end{figure*}

In \autoref{fig:sup_detuning_vs_amplitude_LZ}a, we show the energy of the dressed states of the qubit as a function of the instantaneous detuning $\Delta(t,z)$. The dressed states are separated by an energy gap,  as in the case of the usual Landau-Zener process. However, here both the detuning and amplitude are time- and position-dependent, due to the interplay between the chirping and amplitude of the pulse for different qubit positions. When the pulse has not yet reached the qubit, the energy gap is closed and the qubit is in a specific dressed state. As the pulse interacts with the qubit, the energy gap opens, and the qubit detuning changes, causing the qubit to adiabatically follow the evolution of the dressed state. Specifically, if a qubit, initially prepared in the ground state, is placed at the focal point of the pulse, it is excited and subsequently de-excited. Conversely, if the qubit is not at the focal point, it remains excited after interacting with the pulse, as the energy gap closes while the qubit is excited. 

\autoref{fig:sup_detuning_vs_amplitude_LZ}b and \autoref{fig:sup_detuning_vs_amplitude_LZ}c show the energies of the dressed states as function of time for different experimental conditions, namely for different strengths of the Rabi coupling and different positions of the qubit. Specifically, \autoref{fig:sup_detuning_vs_amplitude_LZ}b shows the energies of the dressed states as a function of time for a qubit placed at the focal point. Here we see that the energy of the dressed state is symmetrical, and has three low energy parts corresponding to the instances in which the qubit is in the ground state (initial), in the excited state (at half the interaction time) and in the ground state (after the interaction with the pulse) again. There exists an optimal value of the Rabi coupling amplitude, such that the coupling is large enough so that the energy gap appears, allowing for the qubit to adiabatically follow one of the dressed states.  As the amplitude of the pulse increases, the rotating-wave approximation becomes less valid, as counter-rotating terms and higher energy levels become significant.  This explains why the described dynamics appears only for a specific pulse amplitude. In \autoref{fig:sup_detuning_vs_amplitude_LZ}c, we show the dressed state energies for different qubit positions. There, for the qubit placed out of the focal point, the eigenenergies are asymmetrical, illustrating the fact that the qubit cannot transition back to the ground state.

\subsection{Impact of reflections}

\begin{figure*}[ht]
    \centering
     \includegraphics[width=1\linewidth]{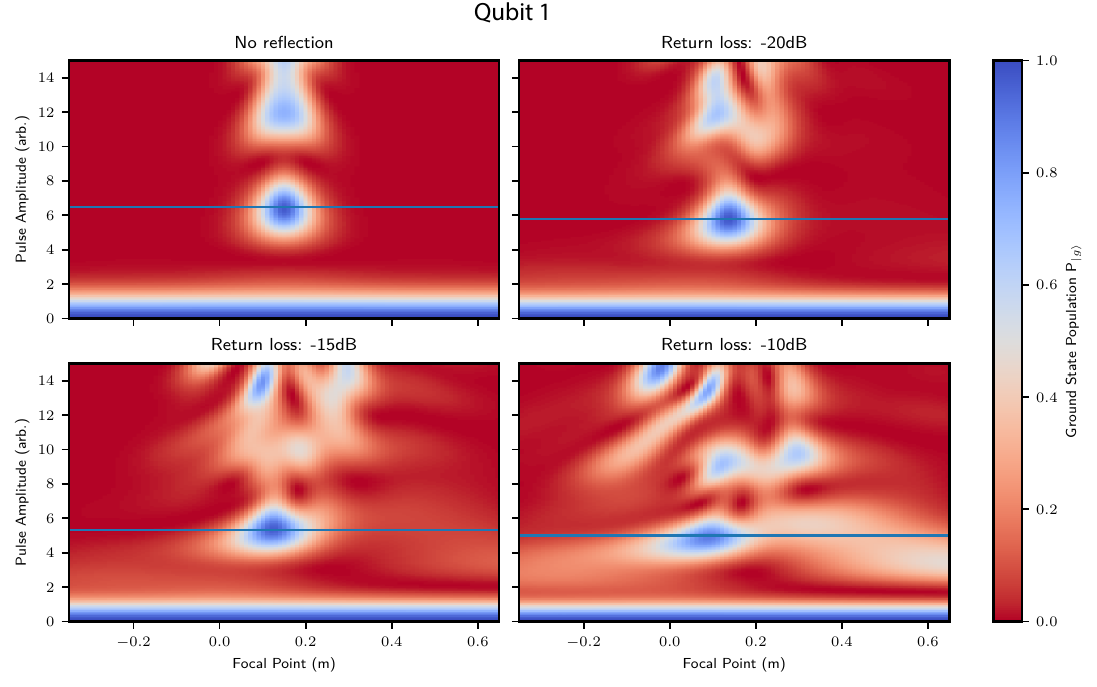}
    \caption{\textbf{Ground state population vs. return loss for Q1.} Simulated ground state population for a transmon at a frequency of 7.2~GHz and an anharmonicity of 420~MHz, after interacting with a pulse with a focal point $z=0~$m and a spot size of 3.5~cm. The different panels show different amplitudes of the parasitic reflection. The reflection considered is constant for all frequencies and the source of the reflection is 10~cm away from the qubit.}
    \label{fig:sup_reflection_impactQ1}
\end{figure*}

\begin{figure*}[ht]
    \centering
     \includegraphics[width=1\linewidth]{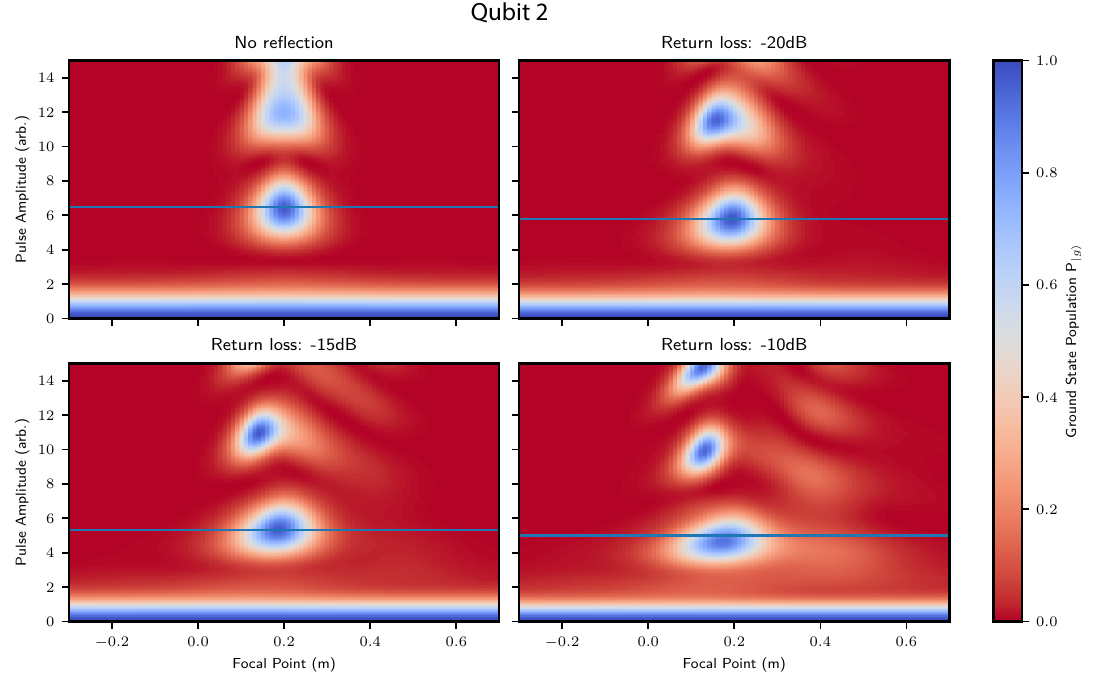}
    \caption{\textbf{Ground state population vs. return loss for Q2.} Simulated ground state population for a transmon at a frequency of 7.2~GHz and an anharmonicity of 420~MHz, after interacting with a pulse with a focal point $z=0~$m and a spot size of 3.5~cm. The different panels show different amplitudes of the parasitic reflection. The reflection considered is constant for all frequencies and the source of the reflection is 5~cm away from the qubit.}
    \label{fig:sup_reflection_impactQ2}
\end{figure*}

The response of the two qubits Q1 and Q2 interacting with a self-compressing pulse is qualitatively different for each qubit, as it can be observed in \autoref{fig:Fig2}. To understand this discrepancy, we perform numerical simulations of the evolution of a transmon qubit interacting with and pulse subject to reflections at the end of the waveguide. In~\autoref{fig:sup_reflection_impactQ1} and \autoref{fig:sup_reflection_impactQ2}, we show the simulated populations for a single reflection 10~cm away from the qubit, for three different amplitudes. The reflection is considered to be point-like, an thus reflects every frequency with the same amplitude and phase.  Additionally, we compare the simulations to the case where no parasitic reflections alter the system. 

\begin{figure*}[ht]
    \centering
     \includegraphics[width=1\linewidth]{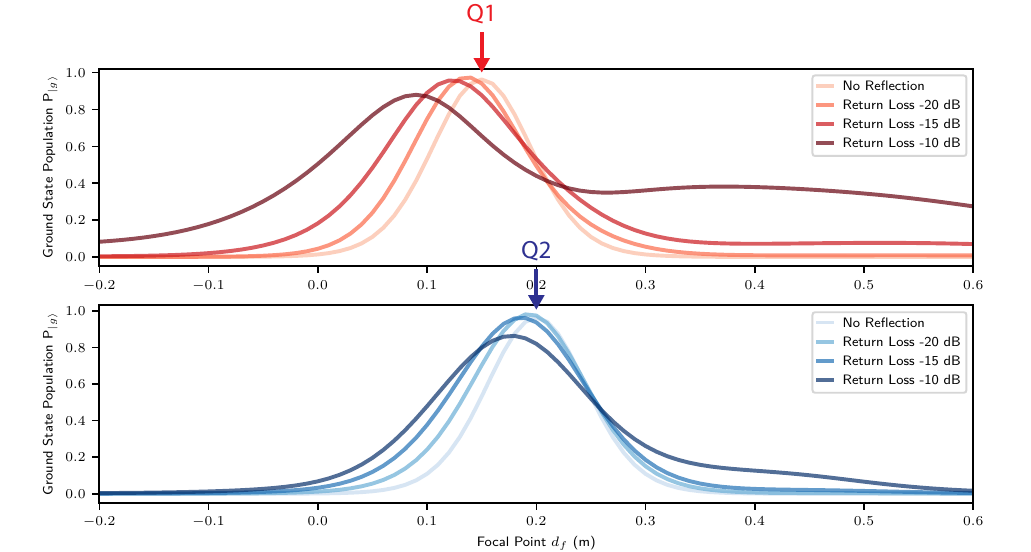}
    \caption{\textbf{Ground state population vs. return loss for Q1 and Q2.} Simulated linecuts of the ground state population for Q1, indicated in~\autoref{fig:sup_reflection_impactQ1}, and for Q2, indicated in~\autoref{fig:sup_reflection_impactQ2}}.
    \label{fig:sup_reflection_linecuts}
\end{figure*}

We observe that, even for moderate reflection amplitudes (with an equivalent return loss of -20~dB or -15~dB), the quantitative behavior of the populations of Q1 and Q2 are significantly different. In \autoref{fig:Fig2}a, the transmission measurement through the waveguide shows several dB of amplitude changes for about \SI{500}{MHz} above the cutoff, which indicates a non-optimal return loss due to impedance mismatches. This effect is due to the rectangular waveguide to coax adapters, which are operated out of the specification regime (the WR90 waveguide is usually operated between 8.20~GHz and 12.40~GHz). Moreover, this main reflection is superimposed with multiple reflections originating from other microwave elements (such as cable or microwave components), which by themselves have specific frequency dependent responses. The superposition of multiple reflections  considerably complicates the dynamic of the qubits, as the exact temporal shape of the pulse affects the qubit state. We observe that Q1 (see \autoref{fig:sup_reflection_impactQ1} and \autoref{fig:sup_reflection_linecuts}) is more affected by reflections than Q2 (\autoref{fig:sup_reflection_impactQ2} and \autoref{fig:sup_reflection_linecuts}), due to the longer round-trip distance that the reflected pulse  travels before returning to Q1.

In order to accurately simulate the system with real experiment reflections, one would need to measure the scattering matrix and compute the impulse response from the qubit for the relevant frequency. One would then be able to do compute the electrical field at the qubit location as a function of the time and obtained it dynamics. It is challenging to do this process accurately at room temperatures, and practically unfeasible at cryogenic temperatures. 
However, for simplicity we instead illustrate the impact of a simple broad band reflection on a qubit. With only a single reflection, we already observe a strong change on the state of the qubit. Frequency dependent reflection leads to a more complicated response, which could better explain the differences the differences in the experimental behavior of the two qubits.

\subsection{Experimental Overview}
\begin{figure*}[ht!]
    \centering
    \includegraphics[width=1\linewidth]{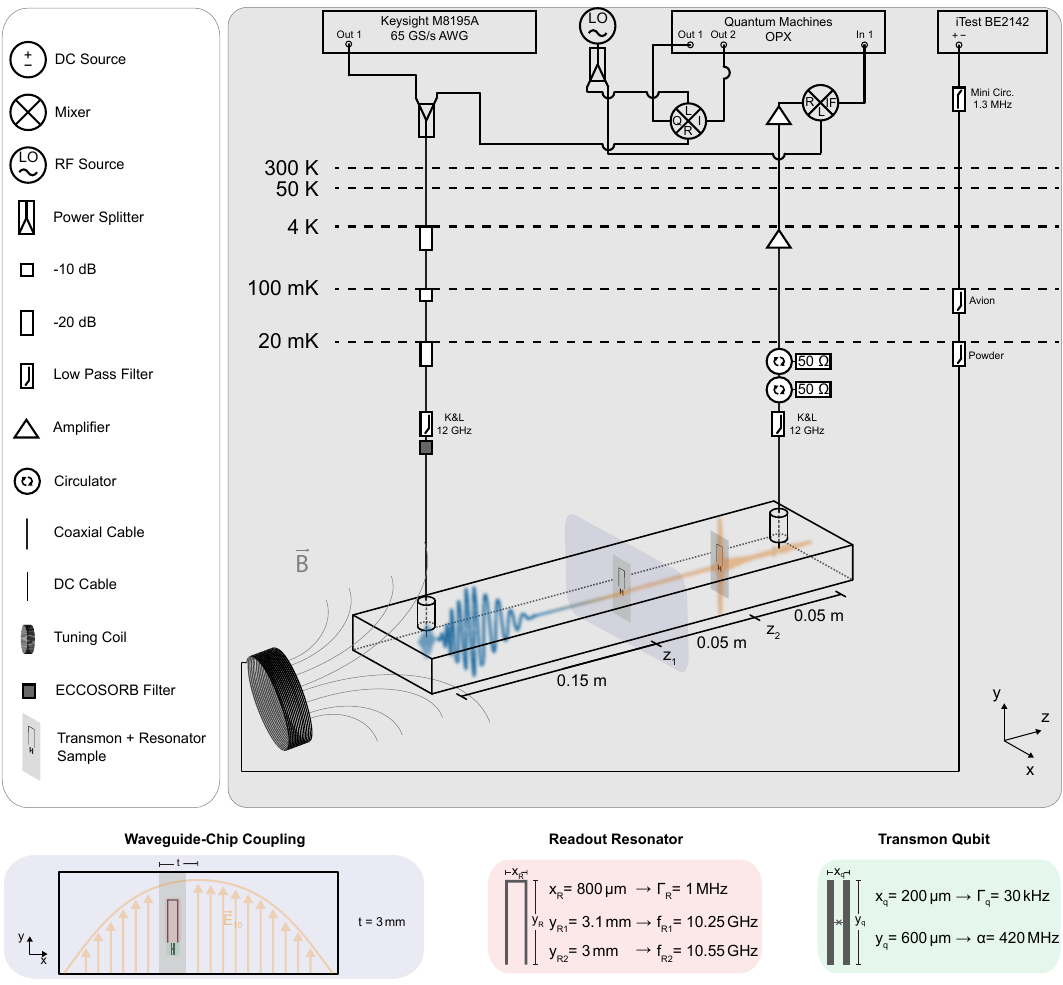}
    \caption{\textbf{Schematic of the experimental wiring with selected components.} Excitation and readout signals are individually generated with the Keysight AWG and Quantum Machines OPX, respectively, and then combined into a single input line that is connected to the waveguide sample. The waveguide output connects to the measurement setup where the readout pulse is down-converted, and then demodulated and digitized by the OPX. The bias coil is located at the output port such that it will supply a magnet field that gradually decays along the z-axis.}
    \label{fig:sup_cd_schematic}
\end{figure*}

The measurements presented in this article are performed in an Oxford Triton Cryofree dilution refrigerator system with a DU7-300 dilution unit, with base temperature of $T = \SI{20}{\milli\kelvin}$. In~\autoref{fig:sup_cd_schematic}, the dashed line at $T = \SI{300}{\kelvin}$ separates the cryostat from the room-temperature electronics. The input coaxial cables are attenuated by \SI{20}{\decibel} at the \SI{4}{\kelvin} stage, followed by by \SI{10}{\decibel} at the \SI{100}{\milli \kelvin} stage and another \SI{20}{\decibel} at the \SI{20}{\milli \kelvin} stage. 
To allow magnetic field tuning of the transmons, the section of the rectangular waveguide containing the qubits is fabricated from oxygen-free high purity copper, while the first half is made from aluminum. The inner dimensions (y,x,z) are $\SI{10.2}{\milli\metre}\times\SI{22.9}{\milli\metre}\times\SI{250}{\milli\metre}$ (input to output coupling port). The commercial adapters at the input-/output of the waveguide are impedance-matched to the \SI{50}{\ohm} impedance of the coaxial cables in order to minimize reflections. The fundamental cutoff frequency $\omega_{c,10}/2\pi=\frac{1}{2a\sqrt{\mu\epsilon}}=\SI{6.546}{\giga\hertz}$ only depends on the longest extension $\rm a=\SI{22.9}{\milli\metre}$ perpendicular to the propagation direction, the vacuum permittivity $\epsilon$ and permeability $\mu$. The next higher mode cutoffs of the waveguide are the TE$_{20}$ mode at $\omega_{c,20}/2\pi=\SI{13.091}{\giga\hertz}$ and TE$_{01}$ mode at $\omega_{c,01}/2\pi=\SI{14.696}{\giga\hertz}$. For frequencies above the cutoff, the electromagnetic field propagates through the hollow core of the waveguide with propagation constant $\beta=\sqrt{k^2-k_c^2}$, defined by the wavevector of the propagating mode $\mathrm{k}=\omega\sqrt{\epsilon\mu}$ and the cutoff wavevector $\mathrm{k_c}=\pi/a$. The phase velocity is then $v_p=\omega/\beta$. The wavelength in the waveguide is $\lambda_g=2\pi/\beta$.

The waveguide input and output are connected to waveguide-to-coaxial adapters that enable to transmit qubit control and readout pulses from the room-temperature control electronics through the waveguide. The waveguide input is connected to a Quantum Machines OPX quantum controller, which triggers an AWG that generates the excitation pulses given in~\autoref{eq:pulse_definition} with a sampling rate of \SI{65}{GS/s}. The readout pulse is a single frequency Gaussian pulse with a standard deviation of $\sigma = \SI{100}{\nano \second}$, while the broadband pulse amplitude and phase is specifically shaped to focus on a given spot resulting in a pulse bandwidth on the order of a few GHz. The tuning-coil is connected to a voltage source (iTest BE2142) that uses a resistor with resistance $R= \SI{1}{\kilo \ohm}$ to convert the voltage into a biasing current that eventually induces a magnetic field at the qubits to change their resonance frequencies. To cancel other external magnetic fields, the waveguide sample is placed into a $\mu$-metal and superconducting shield.

\begin{figure*}[ht!]
    \centering
    \includegraphics[width=1\linewidth]{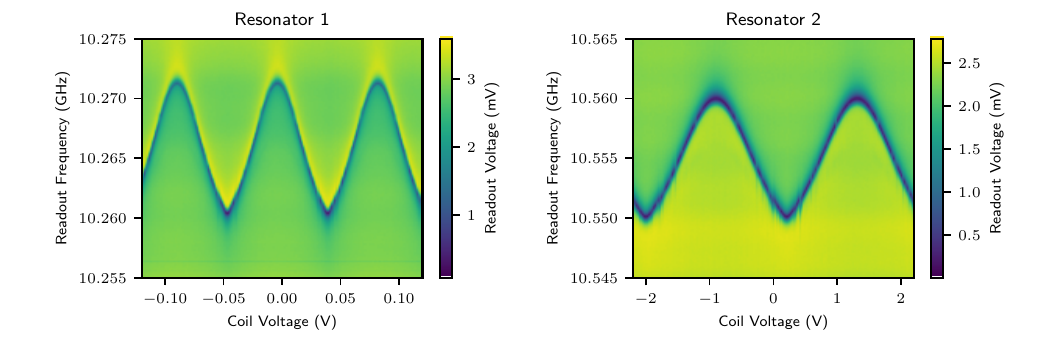}
    \caption{
    \textbf{Resonator Spectroscopy.}
    Transmission magnitude of a readout pulse through the waveguide when changing its frequency and varying the current through a tuning coil. Applying a voltage to a resistor induces a current through a coil that is mounted to the waveguide and changes the flux that threads the SQUID loops of the transmons. This method is used to tune the transmons resonance frequency. Each transmon has a dispersively coupled resonator that tunes according to the detuning between them. As the resonators are coupled directly to the waveguide, their lineshape is observable by measuring the transmission around the resonance frequencies. Resonator 2 needs roughly 20 times the voltage to tune the same amount of oscillations. Following from this we can identify Resonator 1 as the chip closer to the tuning coil.}
    \label{fig:sup_res_vs_flux}
\end{figure*}

The transmon-resonator design is patterned by electron-beam lithography (Raith eLINE Plus \SI{30}{\kilo\volt}) on a bi-layer resist stack (bottom layer: $\SI{1}{\micro\meter}$ of MMA(8.5)MAA EL13, top layer: $\SI{0.2}{\micro\meter}$ of 950 PMMA A4). The substrate are sapphire wafers ($\diameter \SI{50.8}{\milli\meter}$), therefore we sputter a thin layer of gold on top of the PMMA to avoid charging of the sample. After lithography, the gold is etched in a solution of potassium iodide with iodine and water. After developing the resist in a isopropyl alcohol \& water (3:1) solution, two layers of aluminum (\SI{25}{\nano\meter} + \SI{30}{\nano\meter}) are evaporated with a Plassys MEB550S electron-beam evaporator. The junction barrier is formed by a controlled oxidation step before the deposition of the second layer. After liftoff, the samples are laser-diced into individual $\SI{3}{\milli\metre}\times\SI{16}{\milli\metre}$ sized chips and inserted into the waveguide. They are thermalized to the base temperature by a clamp that is attached to the waveguide housing which is screwed onto the cryostat base plate. The transmons are tunable with symmetric Josephson junctions and a maximum frequency of \SI{8.5}{\giga\hertz}. The antenna of the transmon is formed by two rectangular pads of size $A_{\rm pad} = \SI{50}{\micro\metre}\times\SI{600}{\micro\metre}$ (x $\times$ y) separated by a gap of $d_{\rm gap}= \SI{100}{\micro\metre}$ and connected by the wires leading to the junctions. The gradient of the magnetic field supplied by the tuning coil, shown in~\autoref{fig:sup_cd_schematic}, achieves that Q1 and Q2 are subjected to different changes in magnetic field amplitude, when the current through the coil is varied. The distance between the qubits results in a stronger (for Q2) or weaker (Q1) dependence on the applied current, shown in~\autoref{fig:sup_res_vs_flux}. To tune one period of Q1, we need roughly a voltage of \SI{0.1}{\volt}, while we need almost \SI{2}{\volt} for Q2. This enables us to detune the qubits for the single qubit scenario or tune them into resonance for the two qubit experiments. The resonators are patterned above the transmons, allowing to determine the transmon state via dispersive readout~\cite{boissonneaultNonlinearDispersiveRegime2008}.

The samples are embedded in the rectangular waveguide, located at $z_1 = \SI{15}{\centi\meter}$ and $z_2 = \SI{20}{\centi\meter}$, where $z = 0$ corresponds to the input port. The resulting qubit-qubit separation along the photon propagation direction corresponds to half a wavelength $d_z = \SI{5}{\centi\meter} = \lambda/2$ at a frequency of $f = \SI{7.2}{\giga\hertz}$. 
Each chip consists of a transmon qubit~\cite{kochChargeinsensitiveQubitDesign2007b} that is capacitively coupled to a stripline resonator~\cite{zoepflCharacterizationLowLoss2017}. Both, the qubit and resonator antenna, are perpendicular to the propagating electrical field and symmetric along the x-dimension, thus they would not couple to the waveguide if placed in the center. To achieve an electric field gradient across the sample and therefore couple the qubit and resonator to the waveguide, the chip is placed $t=\SI{3}{mm}$ off-center with respect to the maximum electrical field amplitude, shown in the left bottom panel of~\autoref{fig:sup_cd_schematic}. The length of the transmon dipole antenna sets the transmon-waveguide coupling strength to~$\Gamma_{q}/2\pi \approx \SI{30}{\kilo \Hz}$ such that energy relaxation is limited by the coupling to the waveguide~\cite{dalmonteRealizingDipolarSpin2015}. The distance between the resonator arms is used to achieve a resonator-waveguide coupling strength of~$\Gamma_{R1}/2\pi =\Gamma_{R2}/2\pi \approx \SI{1}{MHz}$. The qubit and respective resonator are capacitively coupled with a coupling strength of $g/2\pi \approx \SI{250}{MHz}$. The fundamental resonance frequencies of the readout resonators $\omega_{R1}/2\pi = \SI{10.25}{\giga \hertz}$ and $\omega_{R2}/2\pi = \SI{10.55}{\giga \hertz}$ are visible in the transmission measurement of~\autoref{fig:Fig1}a. The detuning ensures that both transmon qubits can be read out independently.

\subsection{Qubit Coupling}
\begin{figure*}[ht]
    \centering
    \includegraphics[width=1\linewidth]{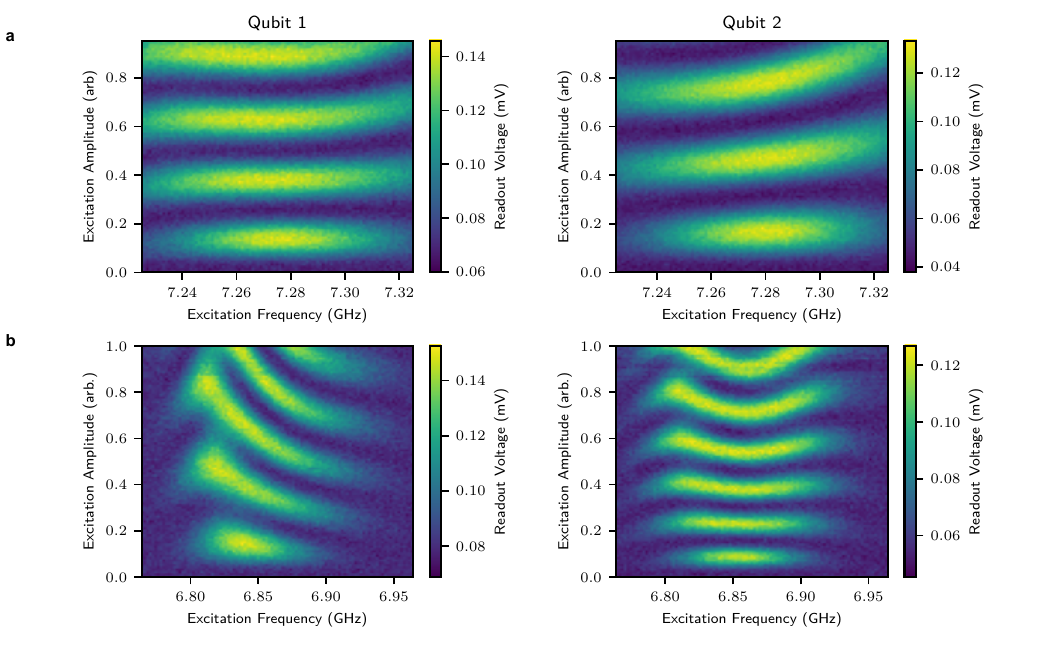}
    \caption{
    \textbf{Rabi Oscillations.}
    \textbf{a} We perform a Rabi experiment, increasing the amplitude of a Gaussian pulse of width $\sigma = \SI{10}{ns}$ while changing its carrier frequency. The occurring Rabi oscillations between the ground state $\ket{g}$ and excited state $\ket{e}$ for Q1 and Q2 are used to calibrate $\pi$-pulses. We attribute the asymmetry different coupling rates between each qubit and the waveguide. \textbf{b} We repeat the experiment around the expected transition frequency of the first excited state $\ket{e}$ and second excited state $\ket{f}$ after initializing the transmon in $\ket{e}$.
     }
    \label{fig:sup_power_rabi_73}
\end{figure*}

\begin{figure*}[ht!]
    \centering
    \includegraphics[width=1\linewidth]{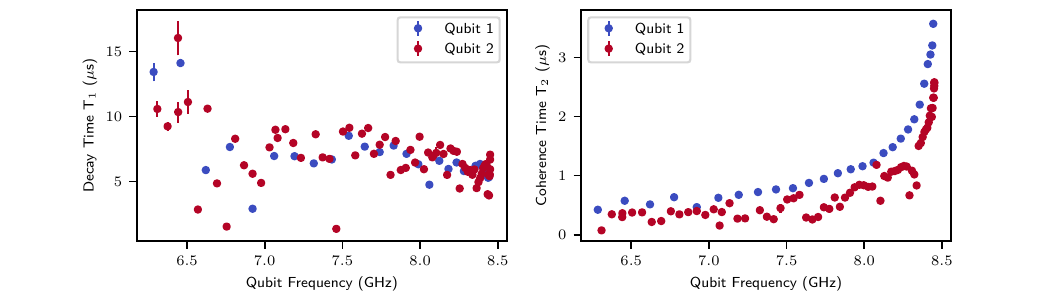}
    \caption{
    \textbf{Qubit Decay.}
    After initializing the qubit to the excited state $\ket{e}$ and waiting a variable delay time, we measure the transmission through the resonator to see the characteristic exponential energy decay with time constant $T_1$. Both qubits decay times are limited by the direct coupling to -- and thus photon emission into -- the waveguide.
    }
    \label{fig:sup_T1_T2}
\end{figure*}

The qubits are characterized by employing conventional Gaussian waveform excitation pulses. In~\autoref{fig:sup_power_rabi_73}, we sweep the amplitude of the pulse while changing its detuning from the qubit frequency. The occurring Rabi oscillations between the ground state $\ket{g}$ and excited state $\ket{e}$ for Q1 and Q2 are used to calibrate $\pi$-pulses. With the calibrated $\pi$-pulse we initialize the qubit in the $\ket{e}$-state and repeat the experiment around the expected transition frequency of the first excited state $\ket{e}$ and second excited state $\ket{f}$. The transition frequency $\omega_{\rm ef}/2\pi$ is $\sim\SI{300}{MHz}$ above the waveguide cutoff frequency. Here, the transmission varies between 70 \% and 100\% in the measured range of~\autoref{fig:sup_power_rabi_73} thus altering the coupling significantly. Even though the chips are designed very similarly, we also observe that Q1 is more strongly coupled to the waveguide. An explanation can be the lower frequency of the readout resonator and thus less Purcell protection by the resonator.
To verify that the qubit decay time $T_1$ is indeed limited by coupling to the waveguide we fit an exponential to the time decay at different frequencies, ranging from the highest possible 
 transon resonance frequency (sweet-spot) to the minimal frequency. The minimal frequency is limited by  attenuation of the pulse below the cutoff. In~\autoref{fig:sup_T1_T2} we show the measured and extracted decay time $T_1$ and coherence time $T_2$, obtained from a Ramsey experiment. The waveguide-limited decay can be experimentally verified by tuning the qubits below the cutoff frequency $f_{\rm c} =\SI{6.56}{\giga \hertz}$. The coupling decreases exponentially when going to lower frequencies than $f_{\rm c}$~\cite{liuQuantumElectrodynamicsPhotonic2017b}, thus we observe an increase in $T_1$. For $T_2$ we observe, the expected decrease when tuning the qubits away from the sweet-spot. Note, that the focusing experiments are performed around \SI{7.2}{\giga\hertz}. In a repetition of the experiment, we can increase $T_2$ by designing the sweet-spot of the transmon flux susceptibility closer to the frequency, where we perform the experiments. 

\subsection{Population Calibration}
\begin{figure*}[ht]
    \centering
    \includegraphics[width=1\linewidth]{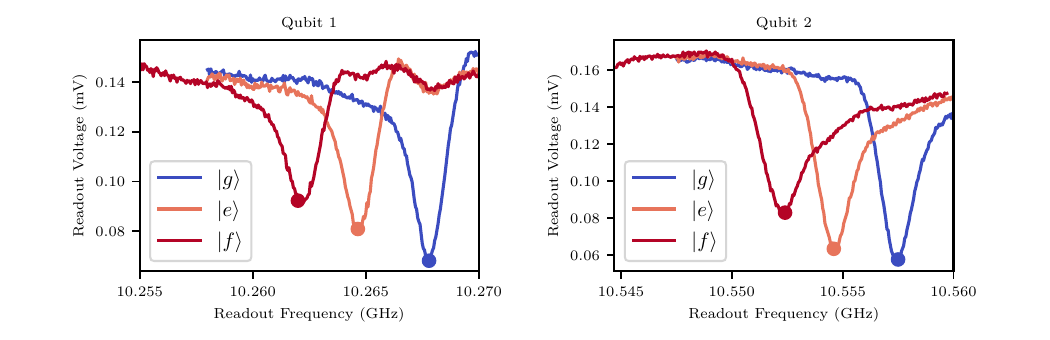}
    \caption{
    \textbf{Dispersive Resonator Shifts.}
    We measure the transmission for frequencies around the resonators when leaving the transmon in the ground state $\ket{g}$ and exciting it to the first excited state $\ket{e}$ or second excited state $\ket{f}$. As expected the resonance frequency of the readout resonator shifts by the dispersive shift $\chi_{ge,Q1}/2\pi = \SI{3.15}{MHz}$ and $\chi_{ge,Q2}/2\pi = \SI{2.65}{MHz}$, when the qubit is prepared in the first excited state $\ket{e}$. When the qubit is prepared in the second excited state $\ket{f}$ the resonator shifts additionally by $\chi_{ef,Q1}/2\pi = \SI{2.9}{MHz}$ and $\chi_{ef,Q2}/2\pi = \SI{2.2}{MHz}$.
    }
    \label{fig:sup_disp_shifts_72}
\end{figure*}
\begin{figure*}[ht]
    \centering
    \includegraphics[width=1\linewidth]{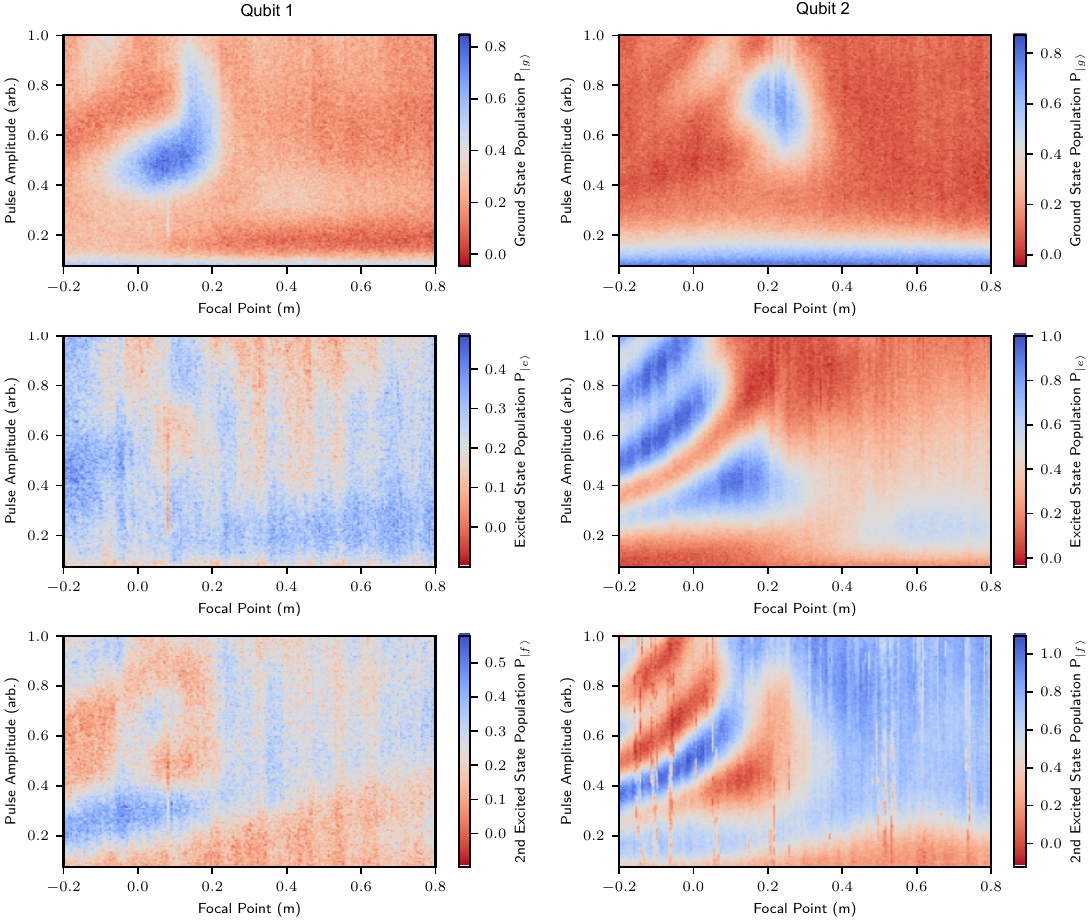}
    \caption{
    \textbf{State Population after Focusing Pulse - One Qubit.} We show the ground state population $P_{\ket{g}}$, and the first $P_{\ket{e}}$ and second excited state populations $P_{\ket{f}}$, of Q1 and Q2 after the interaction with the self-compressing pulse. Here, one qubit has been tuned below the waveguide cutoff where it does not interact. These plots show that most of the population is transferred into the excited states when the focal point matches the physical position of each qubit.    
    }
    \label{fig:sup_focusmaps_gef_72}
\end{figure*}
\begin{figure*}[ht]
    \centering
    \includegraphics[width=1\linewidth]{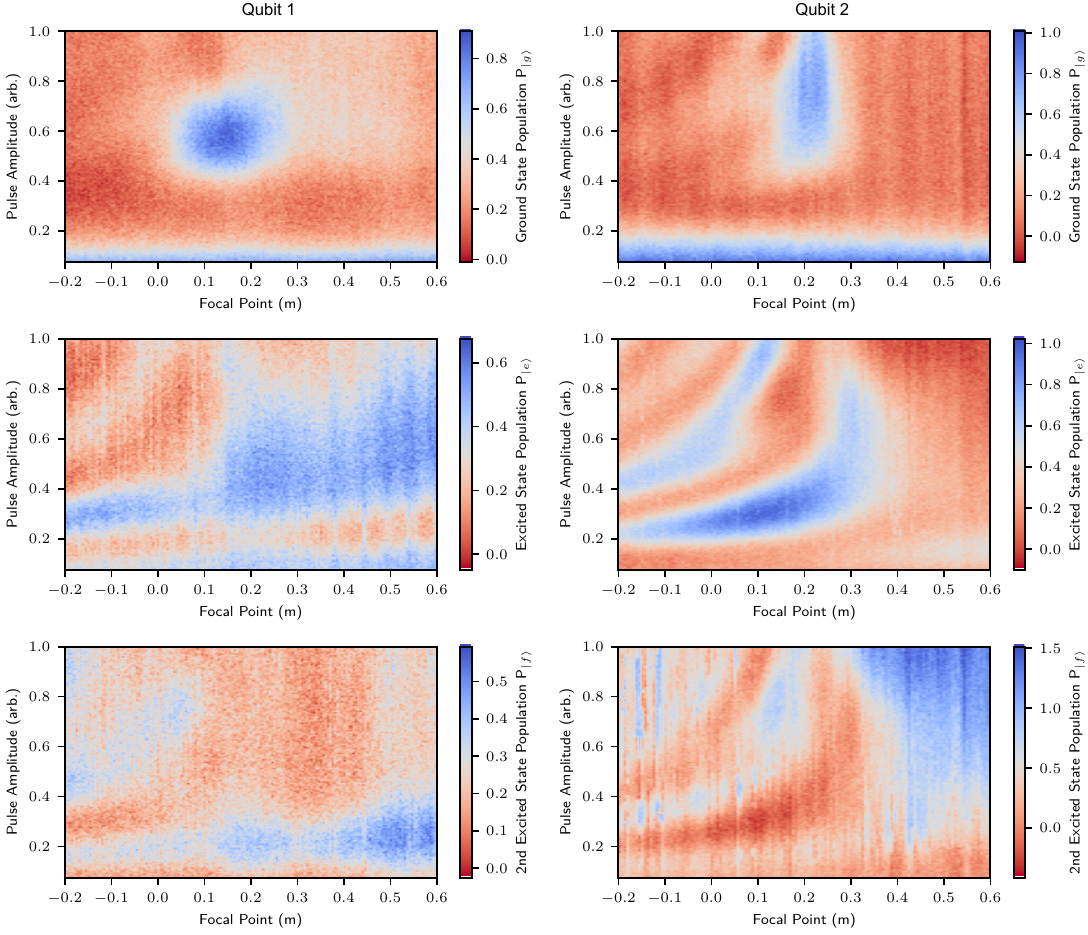}
    \caption{
    \textbf{State Population after Focusing Pulse - Two Qubits.}
    We show the ground state population $P_{\ket{g}}$, and the first $P_{\ket{e}}$ and second $P_{\ket{f}}$ excited state populations, of Q1 and Q2 after interaction with the self-compressing pulse when the qubits are tuned into resonance. These plots show that most of the population is transferred into the excited states when the focal point matches the qubit position.
    }
    \label{fig:sup_focusmaps_gef_73}
\end{figure*}

We measure the transmission through the waveguide QED setup to determine the state of the qubits. The resonators experience a state-dependent dispersive shift $\chi$ and enable us to distinguish between the three lowest energy levels of the transmon qubit. In~\autoref{fig:sup_disp_shifts_72}, we plot the transmission around the resonance frequencies of the resonators, where the minima of the resonant feature varies when we either initially excite the qubits into the first $\ket{e}$ state with a $\pi$-pulse or second excited $\ket{f}$ state with a $\pi$-pulse, followed by a $\pi_{ef}$-pulse or leave the qubit in the ground state $\ket{g}$. The dots indicate the frequency that is used for the readout where the measured voltage corresponds to full population of the respective states. Decreasing amplitude for higher prepared transmon states are mostly due to decay losses during the readout. An increase in transmission amplitude thus corresponds to lower population of the measured state. To calibrate e.g. $P_{\ket{g}} = 0$ we could use the amplitude at the frequency of the resonator when the qubit is in the ground state $f_{\ket{g}}$ when we excite the qubit. However, the focusing pulse depletes the ground state much further than a conventional Gaussian excitation pulse due to the excitation into higher transmon levels, causing a cascaded decay and thus higher transmitted voltage. Using the described calibration this would result in negative state probabilities in~\autoref{fig:Fig2}, ~\autoref{fig:Fig4},~\autoref{fig:sup_focusmaps_gef_72},~\autoref{fig:sup_focusmaps_gef_73}. To calibrate the corresponding probability $P_{\ket{i}} = 0$ for states $i$, we thus use the highest measured voltages when performing the focusing experiment in~\autoref{fig:sup_focusmaps_gef_72} and~\autoref{fig:sup_focusmaps_gef_73}. This underestimates the prepared state fidelities compared to using the calibration coming from a regular Gaussian excitation pulse but ensures positive values in the population measurements. The readout can be improved by utilizing higher order state discrimination and multiplexing of readout signals etc. but is not further considered in the article.

Using the calibration routine described above, we show the population of the states $\ket{g}$, $\ket{e}$ and $\ket{f}$ for only either Q1 or Q2 tuned to $\omega_{ge}/2\pi = \SI{7.2}{\giga \hertz}$ in~\autoref{fig:sup_focusmaps_gef_72} or both qubits tuned to the same frequency in~\ref{fig:sup_focusmaps_gef_73}. The population of states $\ket{e}$ and $\ket{f}$ for all focal points and pulse amplitudes completes the selected linecuts in~\autoref{fig:Fig2}b \&~\ref{fig:Fig4}b. Compared to theory we observe a similar qualitative behavior of distorted features caused by reflections. The distance between Q2 and the output port matches $\lambda/2  = \SI{5}{cm}$. Especially for the excited states, we observe many sudden changes in readout amplitude. These glitches are responsible for the state population exceeding $P_{\ket{i}} = 1$.




\end{document}